\documentclass[12pt,preprint,preprintnumbers,nofootinbib,superscriptaddress,amsmath,amssymb]{revtex4}

\usepackage{graphicx}
\usepackage{dcolumn}
\usepackage{bm}
\usepackage{amssymb}
\usepackage{amsmath}
\usepackage{epsfig}    
\usepackage{color}
\usepackage{slashed}

\begin{document} 

\title{Light charged Higgs boson scenario in 3-Higgs doublet models}

%
\author{A.~G.~Akeroyd}
\email{A.G.Akeroyd@soton.ac.uk}
\affiliation{School of Physics and Astronomy, University of Southampton, Southampton, SO17 1BJ, United Kingdom}
\author{Stefano Moretti}
\email{S.Moretti@soton.ac.uk}
\affiliation{School of Physics and Astronomy, University of Southampton, Southampton, SO17 1BJ, United Kingdom}
\author{Kei Yagyu}
\email{K.Yagyu@soton.ac.uk}
\affiliation{School of Physics and Astronomy, University of Southampton, Southampton, SO17 1BJ, United Kingdom}
\author{Emine Yildirim}
\email{ey1g13@soton.ac.uk}
\affiliation{School of Physics and Astronomy, University of Southampton, Southampton, SO17 1BJ, United Kingdom}

\begin{abstract}

The constraints from the measurements of the $B\to X_s\gamma$ decay rate on the parameter space of 3-Higgs Doublet Models  (3HDMs), where all the 
doublets have non-zero vacuum expectation values, are studied at the next-to-leading order in QCD. 
In order to naturally avoid the presence of flavour changing neutral currents at the tree level, we impose two softly-broken discrete 
$Z_2$ symmetries. This gives rise to five independent types of 3HDMs that differ in their Yukawa couplings.
We show that in all these 3HDMs (including the case of type-II-like Yukawa interactions)  both masses of the two 
charged Higgs bosons $m_{H_1^\pm}$ and $m_{H_2^\pm}$ 
can be smaller than the top mass $m_t$ while complying with the constraints from $B\to X_s\gamma$. 
As an interesting phenomenological consequence, 
the branching ratios of the charged Higgs bosons decay into the $cb$ final states can be as large as $80\%$ 
when their masses are taken to be below $m_t$
in two of the five 3HDMs (named as Type-Y and Type-Z). 
This light charged Higgs boson scenario provides a hallmark 3HDM signature that cannot be realised in $Z_2$ symmetric 2-Higgs doublet models.
We find that in the Type-Y and Type-Z 3HDMs 
the scenario with $90\,{\rm GeV}< m_{H^\pm_1}^{}\,, m_{H^\pm_2}^{}<m_t$ is ruled out
by the direct searches at the LHC, but in the Type-Y 3HDM 
$80~\text{GeV}<m_{H_1^\pm}^{}<90~\text{GeV}$ and $90~\text{GeV}<m_{H_2^\pm}^{}<m_t$ is allowed by $B\to X_s\gamma$ and  
direct searches at LEP2, Tevatron and LHC due to the reduced sensitivity of these searches to the degenerate case $m_{H_1^\pm}\approx m_{W^\pm}$. 
The cases where only one or both charged Higgs bosons are above the top quark mass are also naturally allowed 
in the both Type-Y and Type-Z 3HDMs.

\end{abstract}

\maketitle

\section{Introduction}

After the 7 and 8 TeV runs of the CERN Large Hadron Collider (LHC), 
it has been clarified that a Higgs boson exists with a mass of about 125 GeV and that 
its measured properties -- such as the signal strengths of various production and decay channels -- are consistent with those of the Standard Model 
(SM) Higgs boson \cite{Aad:2012tfa, Chatrchyan:2012xdj}. 
Although this suggests the existence of an isospin doublet scalar field, there remains an open question: i.e., 
how many doublets are there in the actual Higgs sector? 

The existence of a second Higgs doublet is strongly expected when we consider physics Beyond the SM (BSM). 
The most familiar example is the case of supersymmetric extensions of the SM, in which at least two Higgs doublets are required to 
generate all the masses of charged fermions and for  anomaly cancellation~\cite{MSSM}. 
In addition, extra sources of  CP-violation can be obtained from Higgs sectors with a  multi-doublet structure, an ingredient which 
is necessary to realise a successful scenario based on Electro-Weak (EW) baryogenesis~\cite{ewbg1,ewbg2,ewbg3}. 
Furthermore, the second doublet is often introduced in models for neutrino masses~\cite{Zee} and dark matter~\cite{IDM}. 
Therefore, adopting a bottom-up approach while studying the phenomenology of multi-Higgs-doublet models is important in order to access 
BSM physics.

One of the characteristic features of 
extended Higgs models is the appearance of charged Higgs bosons, so that 
their detection can be taken as direct evidence of such structures. 
In particular in multi-doublet models, singly charged Higgs bosons can affect 
various flavour observables such as $B$-meson related processes.
For example,  $B\to X_s\gamma$ data give a lower limit on the mass and Yukawa couplings of charged Higgs bosons. 
In Refs.~\cite{Ciuchini1,Ciuchini2,Borzumati,Gambino}, the Branching Ratio (BR) of $B\to X_s\gamma$ has been calculated at the Next-to-Leading Order (NLO) in QCD
in the context of 2-Higgs Doublet Models (2HDMs) with a softly-broken $Z_2$ symmetry.
In Refs.~\cite{Misiak,Misiak2} the calculation has been extended to Next-to-NLO (NNLO). 
From ~\cite{Misiak2}, the lower limit on the mass of a charged Higgs boson $m_{H^\pm}^{}$ 
is given to be about 480 GeV at 95\% Confidence Level (CL) in the Type-II 2HDM when 
$\tan\beta$, which is the ratio of the Higgs Vacuum Expectation Values (VEVs) of the two doublets, is taken to be larger than 2. 
In contrast, a milder bound from $B\to X_s\gamma$ is extracted in the Type-I 2HDM, e.g., $m_{H^\pm}^{}\gtrsim 100$ and $200$ GeV when 
$\tan\beta = 2.5$ and $2$, respectively~\cite{Misiak}, with the lower bound on  $m_{H^\pm}$ weakening with increasing $\tan\beta$.

It is important to mention here that, in addition to  Type-I and Type-II 2HDMs, 
the Type-X and Type-Y 2HDMs can also be defined depending on the $Z_2$ charge assignment~
\cite{Barger,Grossman,Akeroyd,typeX} and that the same bound on $m_{H^\pm}^{}$ from $B\to X_s\gamma$ as in the Type-II (Type-I) 2HDM is obtained in the Type-Y (Type-X) 2HDM because 
of the identical structure of the quark Yukawa interactions. 
It is then interesting to consider a light charged Higgs boson scenario with $m_{H^\pm} < m_t - m_b$ 
in which $H^\pm$ states can be produced via a top quark decay ($t \to H^\pm b$), a channel which is being searched for at the LHC experiments.
If we consider 2HDMs, such a scenario is allowed in Type-I and Type-X for $\tan\beta\gtrsim 3$ and
it has been shown that the charged Higgs boson mainly decays into $\tau\nu$ in this parameter space  
\cite{Barger, Akeroyd:1994ga, Akeroyd:1995cf,typeX}. 
However, if we consider models with more than two Higgs doublets, one can find charged Higgs bosons decaying copiously into different final states. 

In this paper, we investigate the phenomenology of charged Higgs bosons in 3-Higgs Doublet Models (3HDMs). 
Two softly-broken discrete $Z_2$ symmetries are imposed
in order to realise the Natural Flavour Conserving (NFC) scenario, where 
only one of the three doublets couples to each type of fermion in order to 
avoid Flavour Changing Neutral Currents (FCNCs) at the tree level. 
Under these $Z_2$ symmetries, we define five 
independent types of Yukawa interactions in analogy with the four types of Yukawa interactions in $Z_2$ symmetric 2HDMs. 
In 3HDMs there are two physical charged Higgs bosons (denoted by $H^\pm_1$ and $H^\pm_2$, with $m_{H^\pm_1}<m_{H^\pm_2}$) and
more parameters determine the phenomenology of the charged Higgs sector than in 2HDMs. 
In Refs.~\cite{Grossman, Akeroyd:1994ga, Logan, Akeroyd2}, 
the phenomenology of $H^\pm_1$ in 3HDMs has been studied with decoupled $H^\pm_2$ in terms of 
effective Yukawa couplings for the down-type quark, 
up-type quark and charged lepton, which are expressed by a function of four independent parameters~\cite{Logan} in the framework of NFC. 
It has been shown that 
$H_1^\pm$ can be lighter than the top quark while satisfying constraints from $B\to X_s\gamma$ even for the case with  Type-II like Yukawa couplings. 
Moreover, it was shown in Refs. \cite{Grossman, Akeroyd:1994ga,Akeroyd:1998dt,Akeroyd2} that the decay channel 
$H_1^\pm\to cb$ can have a large BR
(up to $80\%$) in a 3HDM. Although such a value is possible in the Type-Y 2HDM for $m_{H^\pm}< m_t-m_b$,
the constraint $m_{H^\pm}> 480$ GeV from $B\to X_s\gamma$ rules out this scenario. Hence a large BR($H^\pm_1\to cb$) is 
a distinctive signature of 3HDMs.

However, there are some important shortcomings in the 
previous approach where the heavier charged Higgs boson is decoupled from the 
theory. If one takes the decoupling limit of the heavier charged Higgs boson, 
then the mixing angle between the two charged Higgs bosons asymptotically 
approaches zero because of the structure of the charged scalar 
mass matrix. Eventually, this situation makes the predictions 
in 3HDMs identical to those in 2HDMs. In other words, the effective coupling 
approach~\cite{Grossman, Akeroyd:1994ga, Logan, Akeroyd2} is implicitly assuming that a 
cancellation is occurring between the contributions of the two charged Higgs bosons to $B \to X_s \gamma$, 
and the heavier charged Higgs
boson should not be too heavy in order for sufficient cancellation to occur.  
Thus in this paper, we clarify the 3HDM phenomenology with a non-decoupled $H_2^\pm$ and, consequently, the impact
of $H_2^\pm$ on  flavour physics and its typical collider signatures have not been clarified either.
We compute the BR of $B\to X_s\gamma$ at NLO in QCD in 3HDMs  by taking into account both $H_1^\pm$ and $H_2^\pm$ loops, in which 
the dependence of the relevant five independent parameters, i.e., $m_{H_1^\pm}$, $m_{H_2^\pm}$, two ratios of VEVs and a mixing angle between $H_1^\pm$ and $H_2^\pm$, 
is explicitly shown with a fixed type of Yukawa interaction. 
We then discuss the phenomenology of $H_1^\pm$ and $H_2^\pm$ at the LHC in the parameter space allowed by $B\to X_s\gamma$ and 
by the direct searches for  charged Higgs bosons 
via $t \to H^\pm b$ with $H^\pm \to \tau \nu/cs$ at the Tevatron and  LHC as well as pair production
$H^+H^-$ at LEP 2. We draw attention to the fact that current LHC searches for $H^\pm$ do not have sensitivity
to the region $80 \,{\rm GeV} < m_{H^\pm} < 90$ GeV provided that $H^\pm$ has a sizeable branching ratio to $cs$ and/or $cb$,
and LEP2 searches did not rule out
the possibility of a $H^\pm$ in this region.
We then interface these results to the standard hadro-production mode $gg,q\bar q\to t\bar b H^-$ + c.c. discussed in \cite{Miller:1999bm}, where analytical formulae can be found. 

This paper is organised as follows. 
In Sec.~II, we define the 3HDMs. First, we give the Higgs potential under the two softly-broken discrete $Z_2$ symmetries and then we construct the Yukawa Lagrangian.
Five types of Yukawa interactions are also defined.  
In Sec.~III, we discuss the constraints on the parameter space from $B\to X_s\gamma$ and the direct searches for charged Higgs bosons at LEP2, Tevatron and  LHC. 
In Sec.~IV, we investigate the LHC  phenomenology of the charged Higgs bosons. 
Conclusions are given in Sec.~V. 
In Appendix~A, we present the formulae for the mass matrices of the charged, CP-odd and CP-even Higgs bosons. 
In Appendix~B, we summarise all the SM input parameters which are used for the numerical analysis of  this paper.

\begin{table}[t]
\begin{center}
\begin{tabular}{c||c|c|c|c|c|c|c||ccc}
\hline\hline & $\Phi_1$ & $\Phi_2$ & $\Phi_3$ & $u_R^{}$ & $d_R^{}$ & $e_R^{}$ & $Q_L$, $L_L$ &$\Phi_u$ &$\Phi_d$ &$\Phi_e$\\  \hline
Type-I    & $(+,+)$ & $(+,-)$ & $(-,+)$ &$(+,-)$        & $(+,-)$ & $(+,-)$      & $(+,+)$ & $\Phi_2$ & $\Phi_2$ & $\Phi_2$ \\
Type-II  & $(+,+)$ &  $(+,-)$ & $(-,+)$ &$(+,-)$        & $(+,+)$ & $(+,+)$      & $(+,+)$ & $\Phi_2$ & $\Phi_1$ & $\Phi_1$ \\
Type-X   & $(+,+)$ &  $(+,-)$ & $(-,+)$ &$(+,-)$        & $(+,-)$ & $(+,+)$      & $(+,+)$ & $\Phi_2$ & $\Phi_2$ & $\Phi_1$ \\
Type-Y   & $(+,+)$ &  $(+,-)$ & $(-,+)$ &$(+,-)$        & $(+,+)$ & $(+,-)$      & $(+,+)$ & $\Phi_2$ & $\Phi_1$ & $\Phi_2$ \\
Type-Z   & $(+,+)$ &  $(+,-)$ & $(-,+)$ &$(+,-)$        & $(+,+)$ & $(-,+)$      & $(+,+)$ & $\Phi_2$ & $\Phi_1$ & $\Phi_3$ \\
\hline\hline
\end{tabular} 
\end{center}
\caption{Charge assignments under the $Z_2\times \tilde{Z}_2$ symmetry, e.g. ($+,-$) means $Z_2$-even and $\tilde{Z}_2$-odd. } \label{Tab:type}
\end{table}

\section{Models}

We discuss  extensions of the SM Higgs sector with three isospin doublet Higgs fields $\Phi_i$ ($i=1$-3), where 
all the Higgs fields have non-zero VEVs. 
In general, each of these Higgs doublets would couple to all three types of fermions, i.e., up- and down-type quarks and charged leptons. 
However, this structure causes FCNCs at the tree level, as is well known in the general 2HDM without discrete $Z_2$ symmetries. 
The easiest way to avoid FCNCs is to consider a Yukawa Lagrangian where each of the three Higgs doublets couples to at most one of the fermion types,
and such a  Lagrangian takes the following form:
\begin{align}
-{\cal L}_Y = Y_u \bar{Q}_L (i\sigma_2)\Phi_u^* u_R^{} 
+Y_d \bar{Q}_L \Phi_d d_R^{}
+Y_e \bar{L}_L \Phi_e e_R^{} + \text{h.c.}, 
\end{align}
where $\Phi_{u,d,e}$ are either $\Phi_{1}$, $\Phi_2$ or $\Phi_3$. 

\begin{table}[t]
\begin{center}
\begin{tabular}{c||c|c|c||c|c|c}\hline\hline
&\multicolumn{3}{c||}{Factor for $\tilde{H}_1$, $\tilde{A}_1$ and $\tilde{H}_1^\pm$}&\multicolumn{3}{c}{Factor for  $\tilde{H}_2$, $\tilde{A}_2$ and $\tilde{H}_2^\pm$}  \\ \hline
 & $R_{u2}/R_{u1}$ & $R_{d2}/R_{d1}$ & $R_{e2}/R_{e1}$ & $R_{u3}/R_{u1}$ & $R_{d3}/R_{d1}$ & $R_{e3}/R_{e1}$          \\  \hline
Type-I       &$\cot\beta$&$\cot\beta$&$\cot\beta$&0&0&0  \\  \hline
Type-II     &$\cot\beta$&$-\tan\beta$&$-\tan\beta$&0&$-\tan\gamma/\cos\beta$&$-\tan\gamma/\cos\beta$  \\  \hline
Type-X      &$\cot\beta$&$\cot\beta$ &$-\tan\beta$&0&0&$-\tan\gamma/\cos\beta$ \\  \hline
Type-Y      &$\cot\beta$&$-\tan\beta$ &$\cot\beta$&0&$-\tan\gamma/\cos\beta$&0  \\  \hline
Type-Z      &$\cot\beta$&$-\tan\beta$  &$-\tan\beta$&0&$-\tan\gamma/\cos\beta$&$\cot\gamma/\cos\beta$  \\  \hline\hline
\end{tabular} 
\end{center}
\caption{Factors appearing in Eq.~(\ref{yuk2}) for each type of Yukawa interaction. }
\label{ratios}
\end{table}

We can naturally realise the above Lagrangian by imposing two discrete symmetries $Z_2$ and $\tilde{Z}_2$ on the Higgs sector. 
In general, we can also introduce soft-breaking $Z_2$ and $\tilde{Z}_2$ terms in the Higgs potential without losing the key property of 
the absence of FCNCs at  tree level. 
Depending on the charge assignment of the $Z_2$ and $\tilde{Z}_2$ symmetries, 
we can define five independent types of Yukawa interactions\footnote{{We can also define  additional four types by interchanging $\Phi_1 \leftrightarrow \Phi_3$. 
However, these types are physically identical to the last four types given in Tab.~\ref{Tab:type}. }} as listed in Tab.~\ref{Tab:type}. 
We note that the Type-Z corresponds to the Yukawa interaction of the 3HDM discussed in Ref.~\cite{Logan} which 
is named therein as the `democratic 3HDM'.  

The most general Higgs potential under the $SU(2)_L\times U(1)_Y\times Z_2\times \tilde{Z}_2$ symmetry is given by
\begin{align}
V(\Phi_1,\Phi_2,\Phi_3) &= \sum_{i=1}^3 \mu_i^2 \Phi_i^\dagger \Phi_i - (\mu_{12}^2 \Phi_1^\dagger \Phi_2 
+\mu_{13}^2 \Phi_1^\dagger \Phi_3 
+\mu_{23}^2 \Phi_2^\dagger \Phi_3  + \text{h.c.}) \notag\\
 & +\frac{1}{2}\sum_{i=1}^3\lambda_i (\Phi_i^\dagger \Phi_i)^2 
+\rho_1(\Phi_1^\dagger \Phi_1)(\Phi_2^\dagger \Phi_2)
+\rho_2|\Phi_1^\dagger \Phi_2|^2 +\frac{1}{2}[\rho_3(\Phi_1^\dagger \Phi_2)^2 + \text{h.c.} ] \notag\\
&+ \sigma_1(\Phi_1^\dagger\Phi_1)(\Phi_3^\dagger\Phi_3)
 +\sigma_2|\Phi_1^\dagger \Phi_3|^2
+\frac{1}{2}[\sigma_3(\Phi_1^\dagger \Phi_3)^2 + \text{h.c.} ]\notag\\
&+ \kappa_1(\Phi_2^\dagger\Phi_2)(\Phi_3^\dagger\Phi_3)
 +\kappa_2|\Phi_2^\dagger \Phi_3|^2
+\frac{1}{2}[\kappa_3(\Phi_2^\dagger \Phi_3)^2 + \text{h.c.} ], \label{pot}
\end{align}
where the $\mu_{12}^2$, $\mu_{13}^2$ and $\mu_{23}^2$ terms are the soft-breaking terms for $Z_2$ and $\tilde{Z}_2$. 
In general, $\mu_{ij}^2$, $\rho_3$, $\sigma_3$ and $\kappa_3$ are complex parameters but  
throughout the paper we take them to be real for simplicity, thereby avoiding explicit CP violation.
Of the 18 free parameters in the 3HDM scalar potential, two are fixed by the mass of the $W$ boson and
the mass of the discovered neutral Higgs boson. There are theoretical constraints on the 16 remaining parameters 
from requiring stability of the vacuum, absence of charge breaking minima, 
compliance with unitarity of scattering processes etc.
Such constraints are well-known in the 2HDM (e.g. see \cite{Eberhardt:2013uba}) for a recent study)
and have also been discussed for the scalar potential in 3HDMs~\cite{3hdm1,3hdm2,3hdm3,3hdm4,3hdm-uni}. 
We do not impose these constraints because they only rule out certain regions of the 
parameter space of 16 variables, which might not include the region what we are interested in for the phenomenological study focusing on the charged Higgs sector. 
As we will see below, the phenomenology
in the charged Higgs sector depends on only 5 parameters (which we take as unconstrained parameters), 
and we will assume that the freedom in the
other 11 parameters can be used to comply with the above theoretical constraints. To justify this 
approach we note that the analogous constraints on the scalar potential in 2HDMs do not constrain the 
two parameters in the charged Higgs sector ($m_{H^\pm}$ and $\tan\beta$) due to the freedom in the remaining four 
parameters (for the case of a 2HDM with a softly-broken $Z_2$ symmetry). 

The three Higgs doublet fields can be parameterised by 
\begin{align}
\Phi_i = \left[
\begin{array}{cc}
\omega_i^+ \\ 
\frac{1}{\sqrt{2}}(h_i+v_i+iz_i)
\end{array}\right],~~(i=1,...3), 
\end{align}
where the $v_i$'s are the VEVs of the $\Phi_i$'s with the sum rule $\sum_{i}v_i^2\equiv v^2 = 1/(\sqrt{2}G_F) \simeq (246$ GeV$)^2$. 
It is convenient to define the so-called Higgs basis in  3HDMs, 
in which only one of the three doublets contains the VEV $v$ and the Nambu-Goldstone (NG) bosons. 
This can be defined by introducing the orthogonal $3\times 3$ matrix $R$ as 
\begin{align}
\begin{pmatrix}
\Phi_1 \\
\Phi_2 \\
\Phi_3 
\end{pmatrix}=
R
\begin{pmatrix}
\Phi \\
\Psi_1 \\
\Psi_2 
\end{pmatrix}. 
\end{align}
The $R$ matrix is expressed in terms of the three VEVs:
\begin{align}
R &= \begin{pmatrix}
\frac{v_1}{v} &-\frac{v_2v_1}{v_{13}v} &-\frac{v_3}{v_{13}} \\
\frac{v_2}{v} &\frac{v_{13}}{v}                  & 0  \\
\frac{v_3}{v} &-\frac{v_2v_3}{v_{13}v}  &\frac{v_1}{v_{13}}
\end{pmatrix}
=
\begin{pmatrix}
\cos\gamma &0 & -\sin\gamma  \\
0&1&0 \\
\sin\gamma &0 & \cos\gamma   \\
\end{pmatrix}
\begin{pmatrix}
\cos\beta & -\sin\beta & 0\\
\sin\beta &  \cos\beta & 0\\
0&0&1 
\end{pmatrix}\notag\\
&=
\begin{pmatrix}
\cos\beta \cos\gamma & -\sin\beta\cos\gamma & -\sin\gamma\\
\sin\beta &  \cos\beta & 0\\
\cos\beta\sin\gamma&-\sin\beta\sin\gamma&\cos\gamma 
\end{pmatrix}, 
\end{align}
where we introduced the two ratios of the VEVs as follows
\begin{align}
\tan\beta \equiv \frac{v_2}{v_{13}}, \quad 
\tan\gamma \equiv \frac{v_3}{v_1}, ~~\text{with}~~
v_{13} \equiv \sqrt{v_1^2 + v_3^2}. 
\end{align}
Using this notation, each of the VEVs is expressed by 
\begin{align}
v_1 = v_{13}\cos\gamma = v\cos\beta\cos\gamma , \quad 
v_2 =  v\sin\beta,\quad
v_3 = v_{13}\sin\gamma=  v\cos\beta\sin\gamma . 
\end{align}
We note that these definitions differ from those used in Ref.~\cite{Logan}.
In the Higgs basis, the three doublets $\Phi$, $\Psi_1$ and $\Psi_2$ are expressed by 
\begin{align}
\Phi  = 
\begin{pmatrix}
G^+ \\
\frac{v + \tilde{h} + iG^0 }{\sqrt{2}} 
\end{pmatrix}, \quad
\Psi_a  = 
\begin{pmatrix}
\tilde{H}_a^+ \\
\frac{\tilde{H}_a + i\tilde{A}_a }{\sqrt{2}} 
\end{pmatrix} ~~\text{with}~~a=1,2, \label{HB}
\end{align}
where $G^\pm$ and $G^0$ are the NG bosons which are absorbed into the longitudinal components of $W$ and $Z$, respectively. 
In Eq.~(\ref{HB}), two singly-charged states $\tilde{H}_a^\pm$, two CP-odd states $\tilde{A}_a$ and three CP-even states $\tilde{h}$ and $\tilde{H}_a$
are not in general the mass eigenstates. 
The mass eigenstates for the singly-charged ($H_1^\pm$ and $H^\pm_2$) and the CP-odd states ($A_1$ and $A_2$) are defined by
\begin{align}
\begin{pmatrix}
\tilde{H}_1^\pm  \\
\tilde{H}_2^\pm  
\end{pmatrix} = 
\begin{pmatrix}
\cos\theta_C & -\sin\theta_C  \\
\sin\theta_C & \cos\theta_C  
\end{pmatrix}
\begin{pmatrix}
H_1^\pm  \\
H_2^\pm  
\end{pmatrix},~
\begin{pmatrix}
\tilde{A}_1  \\
\tilde{A}_2  
\end{pmatrix} = 
\begin{pmatrix}
\cos\theta_A & -\sin\theta_A  \\
\sin\theta_A & \cos\theta_A  
\end{pmatrix}
\begin{pmatrix}
A_1  \\
A_2
\end{pmatrix}, 
\end{align}
where the mixing angles $\theta_C$ and $\theta_A$ are expressed 
in terms of the mass matrix elements for the singly charged states (${\cal M}_C^2$)
and those for the CP-odd states $({\cal M}_A^2)$ in the Higgs basis (see Appendix~A):
\begin{align}
\tan2\theta_C^{} = \frac{2({\cal M}_C^2)_{12}}{({\cal M}_C^2)_{11}-({\cal M}_C^2)_{22} },\quad 
\tan2\theta_A^{} = \frac{2({\cal M}_A^2)_{12}}{({\cal M}_A^2)_{11}-({\cal M}_A^2)_{22} }. 
\end{align}
The squared mass eigenvalues for the singly-charged and CP-odd Higgs bosons are  given by 
\begin{align}
m_{H_1^\pm}^2 &= ({\cal M}_C^2)_{11}\cos^2\theta_C + ({\cal M}_C^2)_{22}\sin^2\theta_C + ({\cal M}_C^2)_{12}\sin2\theta_C , \\
m_{H_2^\pm}^2 &= ({\cal M}_C^2)_{11}\sin^2\theta_C + ({\cal M}_C^2)_{22}\cos^2\theta_C - ({\cal M}_C^2)_{12}\sin2\theta_C , \\
m_{A_1}^2 &= ({\cal M}_A^2)_{11}\cos^2\theta_A + ({\cal M}_A^2)_{22}\sin^2\theta_A + ({\cal M}_A^2)_{12}\sin2\theta_A , \\
m_{A_2}^2 &= ({\cal M}_A^2)_{11}\sin^2\theta_A + ({\cal M}_A^2)_{22}\cos^2\theta_A - ({\cal M}_A^2)_{12}\sin2\theta_A. 
\end{align}
For the CP-even states, there are three physical states, so that we need to diagonalise the $3\times 3$ mass matrix to obtain the 
mass eigenvalues. The mass eigenstates are defined by introducing the $3\times 3$ orthogonal matrix $R_{H}$ as 
\begin{align}
\begin{pmatrix}
\tilde{h}  \\
\tilde{H}_1\\
\tilde{H}_2 
\end{pmatrix} = 
R_H
\begin{pmatrix}
h  \\
H_1 \\
H_2
\end{pmatrix}. 
\end{align}
Among the three mass eigenstates, one of them must be identified as the discovered Higgs boson at the LHC with a mass of about 125 GeV, which in our case is the  $h$ state. 
The mass matrix for the CP-even states in the Higgs basis $(\tilde{h},\tilde{H}_1,\tilde{H}_2)$ is also given in Appendix~A. 

The Yukawa interaction terms can be expressed in the Higgs basis as 
\begin{align}
&-{\cal L}_Y = \sum_{f=u,d,e}\frac{m_{f^i}}{v}\bar{f}^i\left[\left(\tilde{h} + \frac{R_{f2}}{R_{f1}}\tilde{H}_1 +  \frac{R_{f3}}{R_{f1}}\tilde{H_2} \right) 
-2I_f\left(\frac{R_{f2}}{R_{f1}}\tilde{A}_1 +  \frac{R_{f3}}{R_{f1}}\tilde{A_2} \right) \gamma_5  \right]f^i \notag\\
&+\frac{\sqrt{2}}{v}\left[\bar{u}^j V_{ji} m_{d^i} \left(\frac{R_{d2}}{R_{d1}}\tilde{H}_1^+ +  \frac{R_{d3}}{R_{d1}}\tilde{H}_2^+ \right)P_R  d^i  
-\bar{u}^im_{u^i} V_{ij} \left(\frac{R_{u2}}{R_{u1}}\tilde{H}_1^+ +  \frac{R_{u3}}{R_{u1}}\tilde{H}_2^+ \right)P_L  d^j\right]  + \text{h.c.}\notag\\
&+\frac{\sqrt{2}}{v}\bar{\nu}^i  m_{e^i} \left(\frac{R_{e2}}{R_{e1}}\tilde{H}_1^+ + \frac{R_{e3}}{R_{e1}}\tilde{H}_2^+ \right)P_R  e^i   + \text{h.c.} \label{yuk2}
\end{align}
where $I_f=+1/2\,(-1/2)$ for $f=u\,(d,e)$ and $V_{ij}$ is the Cabibbo-Kobayashi-Maskawa (CKM) matrix element. 
The ratios of the matrix elements $R_{f2}/R_{f1}$ and 
$R_{f3}/R_{f1}$ ($f=u,d,e$) are given in Tab.~\ref{ratios} for each of the five types of Yukawa interactions. 
Here, it is useful to show the correspondence between the $X_a$, $Y_a$ and $Z_a$ couplings used in Refs.~\cite{Grossman, Akeroyd:1994ga, Logan, Akeroyd2} and 
the above couplings. 
When we define these couplings by 
\begin{align}
&-{\cal L}_Y = \frac{\sqrt{2}}{v}\sum_{a=1,2}\left(\bar{u}^j V_{ji} m_{d^i} X_a P_R  d^i  
+\bar{u}^im_{u^i} V_{ij} Y_a P_L  d^j 
+\bar{\nu}^i  m_{e^i} Z_a P_R  e^i \right)H_a^+   + \text{h.c.}, \label{yuk2a}
\end{align}
we find 
\begin{align}
X_1 &= \frac{R_{d2}}{R_{d1}}c_C^{} + \frac{R_{d3}}{R_{d1}}s_C^{},~ 
Y_1 = -\frac{R_{u2}}{R_{u1}}c_C^{} - \frac{R_{u3}}{R_{u1}}s_C^{},~
Z_1 = \frac{R_{e2}}{R_{e1}}c_C^{} + \frac{R_{e3}}{R_{e1}}s_C^{},  \label{xyz1}\\
X_2 &= -\frac{R_{d2}}{R_{d1}}s_C^{} + \frac{R_{d3}}{R_{d1}}c_C^{},~ 
Y_2 = \frac{R_{u2}}{R_{u1}}s_C^{} - \frac{R_{u3}}{R_{u1}}c_C^{},~
Z_2 = -\frac{R_{e2}}{R_{e1}}s_C^{} + \frac{R_{e3}}{R_{e1}}c_C^{},  \label{xyz2}
\end{align}
where $s_C^{}=\sin\theta_C$ and $c_C^{}=\cos\theta_C$. 
In this paper, we particularly focus on the physics related to the charged Higgs bosons $H_1^\pm$ and $H_2^\pm$, for which the number of 
relevant (new physics) parameters is five, namely, 
\begin{align}
m_{H_1^\pm}^{},~m_{H_2^\pm}^{},~\tan\beta,~\tan\gamma,~\text{and}~\theta_C. 
\end{align}
A sixth parameter \cite{Logan}, which is a complex phase $\delta$ of the mass matrix ${\cal M}_C^2$, is 
set to be zero as we have already assumed the CP-invariance of the Higgs sector. 

\begin{figure}[t]
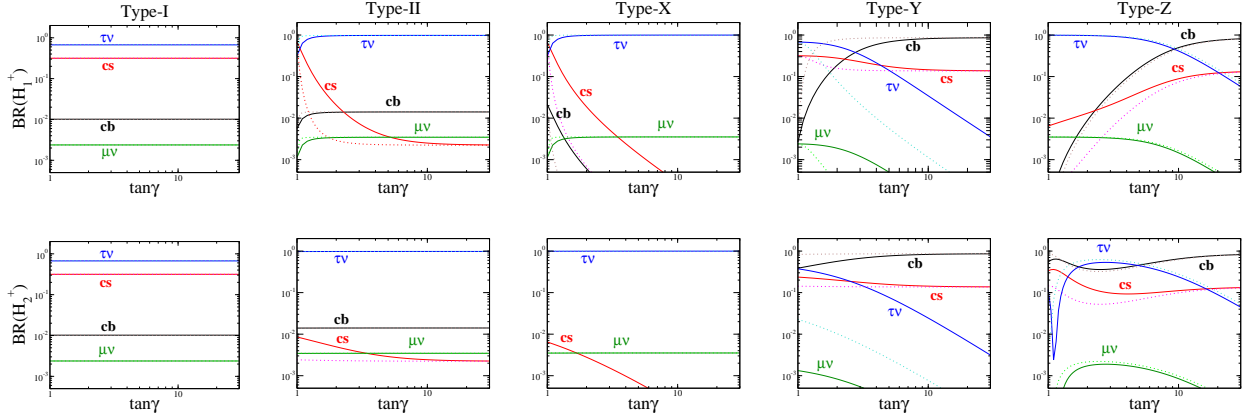

\begin{center}
\includegraphics[width=31mm]{BR_H1p_type1_light.eps}\hspace{4.0mm}
\includegraphics[width=28mm]{BR_H1p_type2_light.eps}\hspace{4.0mm}
\includegraphics[width=28mm]{BR_H1p_typex_light.eps}\hspace{4.0mm}
\includegraphics[width=28mm]{BR_H1p_typey_light.eps}\hspace{4.0mm}
\includegraphics[width=28mm]{BR_H1p_typez_light.eps}\\\vspace{5mm}
\includegraphics[width=31mm]{BR_H2p_type1_light.eps}\hspace{4.0mm}
\includegraphics[width=28mm]{BR_H2p_type2_light.eps}\hspace{4.0mm}
\includegraphics[width=28mm]{BR_H2p_typex_light.eps}\hspace{4.0mm}
\includegraphics[width=28mm]{BR_H2p_typey_light.eps}\hspace{4.0mm}
\includegraphics[width=28mm]{BR_H2p_typez_light.eps}
\caption{Branching ratios of $H_1^\pm$ (upper panels) and $H_2^\pm$ (lower panels) 
as a function of $\tan\gamma$ in the Type-I, II, X, Y and Z 3HDM from the left to right panels.  
We take $m_{H_1^\pm}=100$ GeV, $m_{H_2^\pm}=150$ GeV and $\theta_C= -\pi/4$. 
The value of $\tan\beta$ is taken to be 2 (5) for the solid (dotted) curves. 
}
\label{br1}
\end{center}
\end{figure}

\begin{figure}[t]
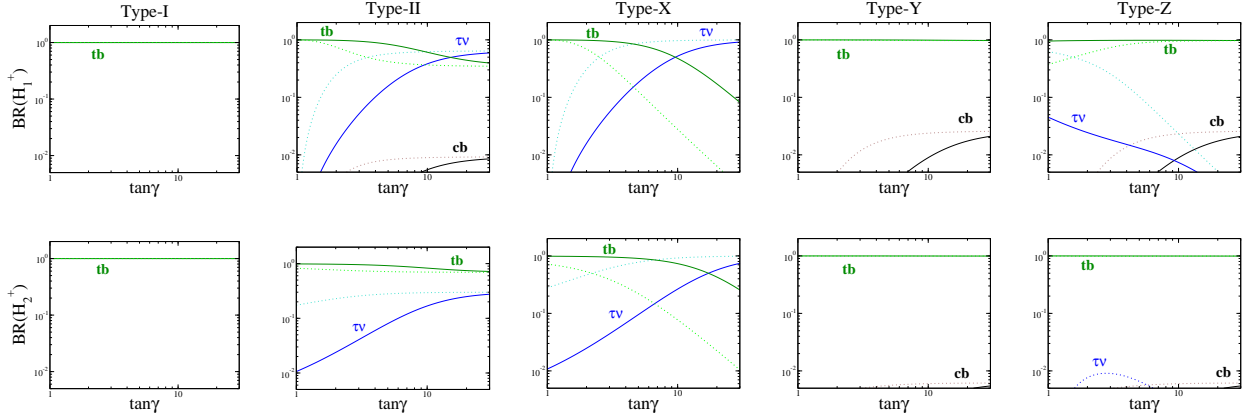

\begin{center}
\includegraphics[width=31mm]{BR_H1p_type1_heavy.eps}\hspace{4.0mm}
\includegraphics[width=28mm]{BR_H1p_type2_heavy.eps}\hspace{4.0mm}
\includegraphics[width=28mm]{BR_H1p_typex_heavy.eps}\hspace{4.0mm}
\includegraphics[width=28mm]{BR_H1p_typey_heavy.eps}\hspace{4.0mm}
\includegraphics[width=28mm]{BR_H1p_typez_heavy.eps}\\\vspace{5mm}
\includegraphics[width=31mm]{BR_H2p_type1_heavy.eps}\hspace{4.0mm}
\includegraphics[width=28mm]{BR_H2p_type2_heavy.eps}\hspace{4.0mm}
\includegraphics[width=28mm]{BR_H2p_typex_heavy.eps}\hspace{4.0mm}
\includegraphics[width=28mm]{BR_H2p_typey_heavy.eps}\hspace{4.0mm}
\includegraphics[width=28mm]{BR_H2p_typez_heavy.eps}\\\vspace{5mm}
\caption{
Branching ratios of $H_1^\pm$ (upper panels) and $H_2^\pm$ (lower panels) 
as a function of $\tan\gamma$ in the Type-I, II, X, Y and Z 3HDM from the left to right panels.  
We take $m_{H_1^\pm}=200$ GeV, $m_{H_2^\pm}=250$ GeV and $\theta_C= -\pi/4$. 
The value of $\tan\beta$ is taken to be 2 (5) for the solid (dotted) curves. 
}
\label{br2}
\end{center}
\end{figure}

We now show the BRs of $H_1^\pm$ and $H_2^\pm$ in the Type-I, Type-II, Type-X, Type-Y and Type-Z 3HDMs.
For simplicity, we take all the masses of extra neutral Higgs bosons ($H_{1,2}$ and $A_{1,2}$) to be larger than those of the charged Higgs bosons and  
take the alignment limit $R_H\to I_{3\times 3}$, 
where the CP-even states in the Higgs basis $\tilde{h}$ and $\tilde{H}_{1,2}$ correspond to the mass eigenstates. 
In this case, the decays of the charged Higgs bosons such as $H_{a}^\pm \to W^{(*)\pm} A_{a}/H_a/h$ ($a=1,2$) do not appear nor do the $H^\pm_2\to H^\pm_1  H_a/h$ ($a=1,2$) ones either\footnote{There is no $H_1^\pm H_2^\mp Z$ vertex at  tree level because  both charged Higgs bosons 
originate from  Higgs fields with identical $SU(2)_L\times U(1)_Y$ charge. Thus, there is no $H_2^\pm \to H_1^\pm Z^{(*)}$ decay at the tree level either. 
We also note that there is no $H_1^\pm H_2^\mp \gamma$ vertex in any models with singly charged Higgs bosons at any order of perturbation 
theory because of a Ward identity. 
}. 
As a result, only the charged Higgs boson decays into fermion pairs need to be considered.
{The leading order expressions for the partial widths of $H^\pm_1$ and  $H^\pm_2$ decaying to fermions are given by:  
\begin{align}
\Gamma(H^\pm_a\to \ell^\pm\nu)&=\frac{G_F m_{H^\pm_a}^{}  }{4\pi\sqrt{2}} m_\ell^2 Z_a^2\left(1-\frac{m_\ell^2}{m_{H_a^\pm}^2} \right)^2, \label{width_tau} \\ 
\Gamma(H^\pm_a\to ud)&=\frac{3G_F |V_{ud}|^2 m_{H^\pm_a}}{4\pi\sqrt{2}}\notag\\
&\times \left[(\bar{m}_d^2X_a^2 + \bar{m}_u^2Y_a^2)\left(1-x_u^a-x_d^a\right)-4\bar{m}_u\bar{m}_d \sqrt{x_u^a x_d^a} X_aY_a \right]\lambda^{1/2}(x_u,x_d),  \label{width_ud}
\end{align}
where $\bar{m}_u$ and $\bar{m}_d$ are the running quark masses evaluated at the scale of the mass of the charged Higgs bosons. 
When calculating the running masses we fix the scale to be 100 GeV, since using the actual charged Higgs boson masses only leads to
small numerical differences from the values obtained when taking the scale to be 100 GeV. 
The parameters $m_u^{\text{pole}}$ and $m_d^{\text{pole}}$ are the pole masses of quarks,
and $x_u^a=(m_u^{\text{pole}})^2/m_{H_a^\pm}^2$ and 
$x_d^a=(m_d^{\text{pole}})^2/m_{H_a^\pm}^2$.
The values of all these quark masses are given in Appendix~B. 
In the above expression, $\lambda$ is the two body phase space function given by $\lambda(x,y)=1+x^2+y^2-2x-2y-2xy$.

In Fig.~\ref{br1}, we show the BRs of $H_1^\pm$ (upper panels)
and $H_2^\pm$ (lower panels) as a function of $\tan\gamma$. 
In these plots, we take $m_{H_1^\pm}=100$ GeV, $m_{H_2^\pm}=150$ GeV and $\theta_C=-\pi/4$. 
The solid and dotted curves show the case for $\tan\beta = 2$ and 5, respectively. 
We see that, in the Type-I, Type-II and Type-X 3HDM,  the decays of the charged Higgs bosons into $\tau\nu$ pairs are dominant. 
In contrast, in the Type-Y and Type-Z 3HDM, the decay into $cb$ can be dominant in the large $\tan\gamma$ region. 
As shown in \cite{Grossman, Akeroyd:1994ga, Akeroyd2}, the parameter space of a large BR of the $cb$ channel corresponds
to $|X_a| \gg |Y_a|, |Z_a|$ (see Eq.~(\ref{width_tau}) and Eq.~(\ref{width_ud})) 
and we have shown that this condition can only be realised in the Type-Y and Type-Z 3HDM.

We show similar plots in Fig.~\ref{br2}, but we here take $m_{H_1^\pm}=200$ GeV, $m_{H_2^\pm}=250$ GeV and $\theta_C=-\pi/4$. 
Except for the Type-X 3HDM, the decay of the charged Higgs bosons into $tb$ is dominant in wide regions of the parameter space. 
In the Type-Y and Type-Z 3HDM, the BR of the $cb$ mode can be at the few percent level in the large $\tan\gamma$ region. 

From these results, it can be seen that the charged Higgs boson decay into $cb$ can be important in 
the Type-Y and Type-Z 3HDMs, especially when the charged Higgs boson masses are below the top mass.
We would finally like to emphasise that the $H_{1,2}^\pm \to cb$ decay can be a useful tool to distinguish 3HDMs from 2HDMs because 
of the following reason. 
In practice, in the 2HDM with a softly-broken $Z_2$ symmetry, 
the $H^\pm \to cb$ decay can be dominant when $m_{H^\pm}^{}<m_t - m_b$ and $\tan\beta \gtrsim 3$ for the Type-Y case~\cite{Akeroyd:1994ga, typeX}. 
However, such a light charged Higgs boson is excluded by the $B\to X_s\gamma$ data. 
In our model, 
the constraint from $B\to X_s\gamma$ is instead avoidable using a  cancellation between the contributions from the loops involving 
$H_1^\pm$ and $H_2^\pm$ as we will 
clarify  in the next section.

\section{Constraints from $B\to X_s\gamma$ and direct searches}

In this section, we first discuss the constraints on the parameter space of
the five types of 3HDMs from measurements of $B \to X_s\gamma$. Then we move on to consider constraints from direct
searches for $H_1^\pm$ and $H_2^\pm$ at colliders.

\subsection{Flavour sector limits}

We calculate the branching fraction of the radiative $B \to X_s\gamma$ decay process at  NLO in QCD. 
In our model, in addition to the  $W^\pm$ boson loop contribution, $H_1^\pm$ and $H_2^\pm$ also contribute to this process at the same perturbative level. 
The decay rate of $B \to X_s\gamma$ can be written as a sum of the following three parts: i)
the $b$ quark decay process $b\to s\gamma$ $(\Gamma_{b\to s\gamma})$; ii) the gluon bremsstrahlung process $b\to s\gamma g$ 
$(\Gamma_{b\to s\gamma g})$; iii) non-perturbative effects due to the mesonic processes $(\Gamma_{\text{non-pert.}})$. 
Thus, 
 \begin{align}
\Gamma(B\to X_s\gamma) = \Gamma_{b \to s\gamma} + \Gamma_{b \to s\gamma g} + \Gamma_{\text{non-pert.}}. 
\end{align}
The first and the second contribution depend on the new physics parameters such as the charged Higgs boson masses and  their couplings to quarks, while 
the third contribution does not. 
The decay rates $\Gamma_{b \to s\gamma}$ and $\Gamma_{b \to s\gamma g}$ are calculated using
 the Wilson coefficients at a scale $\mu$ \cite{Hewett:1994bd}:
\begin{align}
C_i^{\text{eff}}(\mu,m_{H_1^\pm},m_{H_2^\pm}) = C_{i,\text{SM}}^{\text{eff}}(\mu)
+\sum_{a=1,2}\left[(X_aY_a^*)C_{i,XY}^{\text{eff}}(\mu,m_{H_a^\pm})+|Y_a|^2C_{i,YY}^{\text{eff}}(\mu,m_{H_a^\pm})\right],   \label{wc}
\end{align}
where $i=1,\dots 8 $ while $X_a$ and $Y_a$ ($a=1,2$) are given in Eqs.~(\ref{xyz1}) and (\ref{xyz2}).
We note that the results of the 2HDMs can be reproduced by taking the limit of $\theta_C \to 0$ or $m_{H_{2}^\pm} \to m_{H_{1}^\pm}$. 
The latter holds due to the sum rule:
\begin{align}
\sum_{a=1,2} X_aY_a = X_1Y_1\Big|_{\theta_C=0},\quad  \sum_{a=1,2} |Y_a|^2 = |Y_1|^2\Big|_{\theta_C=0}. 
\end{align}
The structure of the quark Yukawa couplings are the same in Type-II, Type-Y and Type-Z 3HDMs and so 
the same bound from $B\to X_s\gamma$ applies equally to these three models. Likewise, 
the bound from $B\to X_s\gamma$ applies equally to the Type-I and Type-X 3HDMs.
To obtain $\Gamma_{b\to s\gamma}$ and $\Gamma_{b\to s\gamma g}$, we set the scale $\mu$ appearing in Eq.~(\ref{wc}) 
to be the bottom quark mass scale $\mu_b$. 
All the Wilson coefficients that are calculated at the matching scale $\mu = \mu_W^{}$ have to be evaluated at $\mu_b$ by solving the renormalisation group equations. 
In Ref.~\cite{Borzumati}, all the relevant Wilson coefficients at $\mu=\mu_b$ at LO and NLO are given in terms of 
those at $\mu=\mu_W^{}$ and we adopt them for our numerical evaluations. 
Using the decay rate and BR of the semi-leptonic decay of the $B$ meson, we can express 
the BR of the $B\to X_s\gamma$ process as
\begin{align}
{\rm BR}(B\to X_s\gamma) &= \frac{\Gamma(B\to X_s\gamma)}{\Gamma(B\to X_c\ell\nu)}{\rm BR}(B\to X_c \ell\nu). 
\end{align}
The measured value of the BR is given~\cite{HFAG} as 
\begin{align}
&{\rm BR}(B \to X_s\gamma) = (3.43 \pm 0.22)\times 10^{-4}. 
\end{align}
All the SM input parameters for the numerical calculations are listed in Appendix~B, and we take $\mu_b = m_b^{\text{pole}}$ and $\mu_W = m_{W^\pm}$.

\begin{figure}[t]
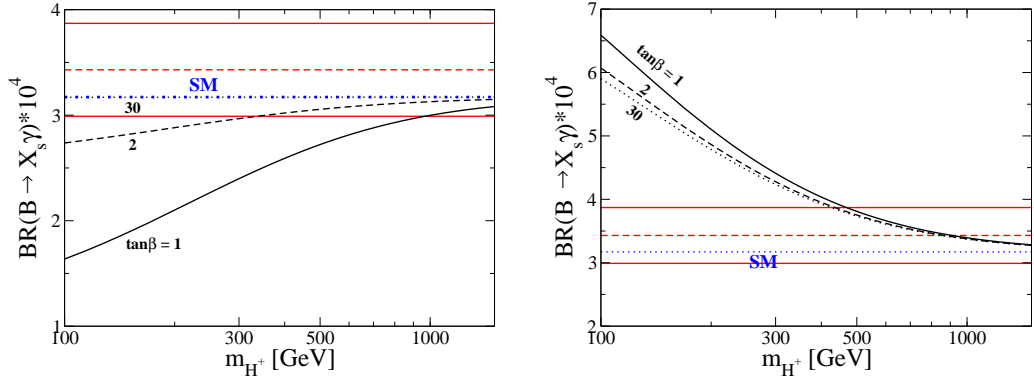

\begin{center}
\includegraphics[width=65mm]{2HDM_type1.eps}\hspace{5mm}
\includegraphics[width=65mm]{2HDM_type2.eps}
\caption{Predictions for the branching ratio of the $B\to X_s\gamma$ process in the Type-I (left) and Type-II 2HDM (right)
as a function of $m_{H^\pm}^{}$. We take $\tan\beta=1$ (solid), 2 (dashed) and 30 (dotted). 
The red solid (dashed) lines are the $2\sigma$ allowed region (central value) of the experimental result. 
The blue dotted line is the SM prediction. 
}
\label{fig1}
\end{center}
\end{figure}

In order to compare the predictions in  2HDMs with those in  3HDMs, we first show the results in the former 
case where we have only one pair of charged Higgs bosons $H^\pm$.  
In Fig.~\ref{fig1}, we show the prediction of the BR of $B\to X_s\gamma$ in the 
Type-I (left) and Type-II (right) 2HDM as a function of the mass of the charged Higgs boson $m_{H^\pm}^{}$. 
The SM prediction is indicated as the blue dotted line. 
The $2\sigma$ bounds from the experimental data are shown as  horizontal red solid lines. 
The black curves show the results in  2HDMs with  several fixed values of $\tan\beta$. 
We can see that the $H^\pm$ contribution interferes  destructively (constructively) with the SM contribution in the Type-I (Type-II) 2HDM. 
In addition, in Type-I, when we take a large $\tan\beta$ value, the $H^\pm$ contribution becomes quite small because  all the $H^\pm$ couplings to quarks are 
proportional to $\cot\beta$. 
In contrast,  in the Type-II case, even if we take a large $\tan\beta$ value, the $H^\pm$ loop effect does not vanish. 
This can be understood by noting that the coupling product $X_1Y_1^*$ with $\theta_C \to 0$ appearing in the Wilson coefficient is equal to 
unity in the Type-II 2HDM. 
Consequently, in  Type-I,  a severe lower limit on $m_{H^\pm}$ is only obtained for small values of $\tan\beta$, 
with the bound being about $m_{H^\pm}> 1$ TeV with $\tan\beta = 1$. 
In Type-II, when $\tan\beta \gtrsim 2$, we obtain $m_{H^\pm} \gtrsim 450$ GeV independently of 
$\tan\beta$\footnote{In Ref.~\cite{Misiak2}, 
the calculation at NNLO in QCD has been evaluated in the Type-II 2HDM and 
the slightly more stringent  limit $m_{H^\pm}\gtrsim 480$ GeV has been derived at 95\% CL.}. 

\begin{figure}[t]
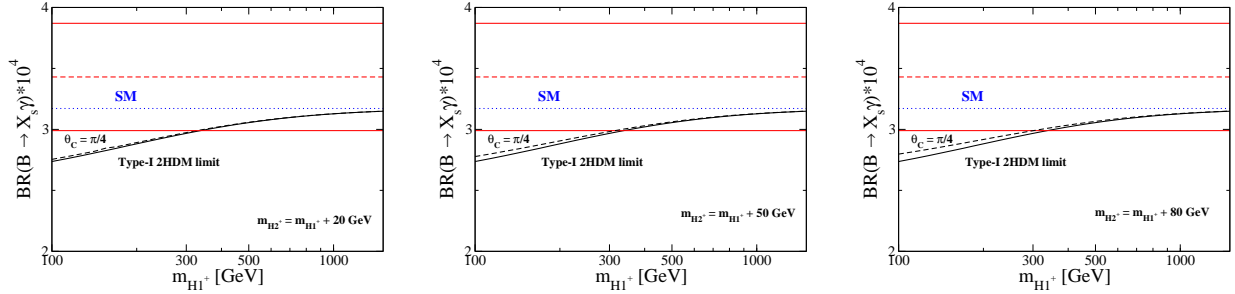

\begin{center}
\includegraphics[width=50mm]{3HDM_type1_delm20.eps}\hspace{5mm}
\includegraphics[width=50mm]{3HDM_type1_delm50.eps}\hspace{5mm}
\includegraphics[width=50mm]{3HDM_type1_delm80.eps}\hspace{5mm}
\caption{Prediction of the branching ratio of $B\to X_s\gamma$ in the Type-I 3HDM (black dashed curve) with $\tan\beta=2$ and $\theta_C = \pi/4$
as a function of $m_{H_1^\pm}^{}$.   
As a comparison, we also show the results in the Type-I 2HDM with $\tan\beta = 2$ as the black solid curve. 
The left, centre and right panels show the case for $m_{H_2^\pm}^{}-m_{H_1^\pm}^{}=20$, 50 and 80 GeV, respectively. 
The same results are obtained in the Type-X 3HDM. }
\label{fig2}
\end{center}
\end{figure}

Next, we show the numerical results of ${\rm BR}(B\to X_s\gamma)$ in 3HDMs. 
In Fig.~\ref{fig2}, the $m_{H_1^\pm}^{}$ dependence of ${\rm BR}(B\to X_s\gamma)$ in the Type-I 3HDM is shown as the black dashed curve. 
The prediction in the Type-I 2HDM is also shown as the solid curve for comparison. 
In these plots, we take $\tan\beta = 2$ and $\theta_C = \pi/4$. 
The mass difference $m_{H_2^\pm}-m_{H_1^\pm}$ is taken to be 20 (left panel), 50 (centre panel) and 80 GeV (right panel). 
We can see that the difference between the prediction in the Type-I 3HDM and the Type-I 2HDM
becomes slightly bigger as the mass difference $m_{H_2^\pm}-m_{H_1^\pm}$ increases, but,  
even for the case of $m_{H_2^\pm}-m_{H_1^\pm}=80$ GeV, these two results are almost the same. 
We note that ${\rm BR}(B\to X_s\gamma)$ does not depend on $\tan\gamma$ in the Type-I 3HDM, because
$\tan\gamma$ is not entering the quark Yukawa couplings as shown in Tab.~\ref{ratios}. 
We also note that the prediction in the Type-I 3HDM does not depend on the sign of $\theta_C$. 

\begin{figure}[t]
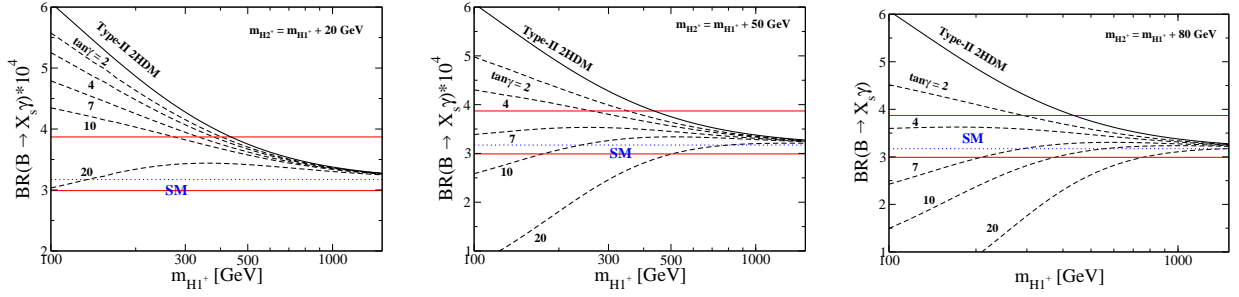

\begin{center}
\includegraphics[width=50mm]{3HDM_type2A_delm20.eps}\hspace{5mm}
\includegraphics[width=50mm]{3HDM_type2A_delm50.eps}\hspace{5mm}
\includegraphics[width=50mm]{3HDM_type2A_delm80.eps}\hspace{5mm}
\caption{Prediction of the branching ratio of $B\to X_s\gamma$ in the Type-II 3HDM with several values of $\tan\gamma$.  
We take $\tan\beta=2$ and $\theta_C=-\pi/4$. 
As a comparison, we also show the results in the Type-II 2HDM with $\tan\beta = 2$ as the black solid curve. 
The left, centre and right panels show the case for $m_{H_2^\pm}^{}-m_{H_1^\pm}^{}=20$, 50 and 80 GeV, respectively. 
The same results are obtained in the Type-Y and Type-Z 3HDM. 
}
\label{fig3}
\end{center}
\end{figure}

In Fig.~\ref{fig3}, 
the $m_{H_1^\pm}^{}$ dependence of ${\rm BR}(B\to X_s\gamma)$ in the Type-II 3HDM for several fixed values of $\tan\gamma$ is shown as  black dashed curves. 
The prediction in the Type-II 2HDM is also shown as the solid curve for comparison. 
In these plots, we take $\tan\beta = 2$ and $\theta_C = -\pi/4$. 
The mass difference $m_{H_2^\pm}-m_{H_1^\pm}$ is taken to be 20, 50 and 80 GeV in the left, centre and right panel, respectively. 
It is clear that the prediction in the 3HDM becomes smaller when we take a larger value of $\tan\gamma$. 
This tendency becomes more evident with  larger mass differences. 
As a result, we can find that cases where both charged Higgs boson masses of ${\cal O}(100)$ GeV are allowed by taking 
 appropriate values for $\tan\gamma$ and  their mass difference. 

We here comment on the constraint on the parameter space from the other observables in flavour physics according to Ref.~\cite{Logan}. 
We note that the constraints discussed in Ref.~\cite{Logan} are based on a 3HDM with  $H_2^\pm$ decoupled, so that 
we cannot simply apply them to our case.  
{In the following, we apply these constraints to get the limit on each of the couplings for $H_1^\pm$ and $H_2^\pm$, which means that we do not 
take into account the interference effect of the two charged Higgs boson contributions.  }

From $R_b$ measured from the $Z \to b\bar{b}$ decay, we obtain 
\begin{align}
|Y_a| \leq 0.72 + 0.24\left(\frac{m_{H_1^\pm}}{100~\text{GeV}}\right)~~\text{at 95\% CL}, 
\end{align}
under $|X_a|<50$ ($a=1,2$). 
This can be easily avoided by taking $\tan\beta \gtrsim 1$ for $m_{H_1^\pm}=100$ GeV. 
The bound on the charged lepton coupling $Z_1$ is obtained from the leptonic $\tau$ decay as
\begin{align}
Z_a \leq 40\left(\frac{m_{H_1^\pm}}{100~\text{GeV}}\right)~~\text{at 95\% CL}. 
\end{align}
This corresponds to the bound on $\tan\gamma \lesssim 32(15)$ for $\tan\beta=2(5)$ with $m_{H_1^\pm}=100$ GeV and $\theta_C = -\pi/4$ 
in the Type-II and Type-X 3HDMs. 
For the other types, this does not set an upper limit on $\tan\gamma$ unless we take $\tan\beta \gg 1$ and/or $m_{H_1^\pm}\ll 100$ GeV. 
Finally, from the measurement of $B\to \tau\nu$, we obtain 
\begin{align}
|X_a Z_a| \leq 1080\left(\frac{m_{H_1^\pm}}{100~\text{GeV}}\right)^2~~\text{at 95\% CL}. 
\end{align}
This gives an important constraint on the parameters only in the Type-II 3HDM, because both $X_1$ and $Z_1$ are enhanced by 
increasing $\tan\beta$ and $\tan\gamma$. For example, $\tan\gamma\gtrsim 22(11)$ is excluded when $\tan\beta=2(5)$, $m_{H_1^\pm}=100$ GeV and $\theta_C=-\pi/4$. 
{We checked that all these above constraints are satisfied in the numerical analysis presented in the succeeding sections. }

\subsection{Collider limits}

As discussed in the previous subsection, in 3HDMs, we can take the charged Higgs boson masses to be ${\cal O}(100)$ GeV 
without conflict with the $B\to X_s \gamma$ data. 
In this subsection we discuss this scenario at the LHC as a hallmark manifestation of a 3HDM,
particularly for the Type-Y and Type-Z cases, because 
the characteristic decay of the charged Higgs bosons $H_{1,2}^\pm \to cb$ can be dominant. 

When we consider the case for $m_{H_{1,2}^\pm}^{}<m_t-m_b$, we need to take into account the constraints from  direct searches for $H^\pm$ states from the 
top quark decay $t\to H^\pm b$ at the LHC. 
In Ref.~\cite{taunu}, ATLAS carried out a search for the decay $H^\pm \to \tau\nu$ using the data taken with 8 TeV of collision 
energy and 19.5 fb$^{-1}$ of integrated luminosity. 
From the non-observation of an excess above the SM prediction, the 95\% CL lower limit on ${\rm BR}(t \to H^\pm b)\times {\rm BR}(H^\pm \to \tau^\pm\nu)$
has been given to be between 0.23\% and 1.3\% in the range  $80$ GeV $<m_{H^\pm}<160$ GeV. Similar limits are derived in the
CMS search in \cite{Khachatryan:2015qxa}.
The search for $H^\pm$ with decay into $cs$ has also been performed in \cite{hadronic} by CMS 
using 8 TeV data and 19.7 fb$^{-1}$ of  integrated luminosity. 
The 95\% CL lower limit on ${\rm BR}(t \to H^\pm b)\times {\rm BR}(H^\pm \to cs)$
has been given to be between 1.2\% and 6.5\% in the range  $90$ GeV $<m_{H^\pm}<160$ GeV. Similar constraints are
obtained from the ATLAS  search for $H^\pm\to cs$ in \cite{Aad:2013hla}. We note that
there is a local excess of $2.4\sigma$ around $m_{H^\pm}=150$ GeV in the CMS search in \cite{hadronic}, with a
best-fit branching fraction of $t\to H^\pm b=1.2\pm 0.2\%$, assuming BR$(H^\pm\to cs)=100\%$.

In order to estimate the bound from these LHC direct searches in our 3HDMs, we require the following conditions as the strongest bound of which meaning is explained below:
\begin{align}
&\sum_{a=1,2}{\rm BR}(t\to H_a^\pm b) \times {\rm BR}(H_a^\pm \to \tau^\pm\nu) < 0.23\%, \label{leptonic} \\
&\sum_{a=1,2}{\rm BR}(t\to H_a^\pm b) \times [{\rm BR}(H_a^\pm \to cs)+{\rm BR}(H_a^\pm \to cb)] < 1.2\%, \label{had}
\end{align}
where these constraints can be applied to the case of $90~\text{GeV}< m_{H_{1,2}^\pm} <160$ GeV. 
Regarding the second equation, we include the $cb$ mode because no flavour tagging was employed in Ref.~\cite{hadronic}, 
see also \cite{Akeroyd2, Akeroyd:2014cma}.
We note that the two charged Higgs boson contributions should not be summed if the mass difference between $H_1^\pm$ and $H_2^\pm$
is taken to be larger than the detector resolution. If we do not sum these two contributions, then 
we should get a milder bound than that obtained from Eqs.~(\ref{leptonic}) and (\ref{had}). 
Thus, the meaning of ``strongest bound'' is choosing the strongest limit on the product of two branching fractions 
(the top decay and the charged Higgs boson decay) 
in the given mass range and summing two charged Higgs boson contributions.

\begin{figure}[t]
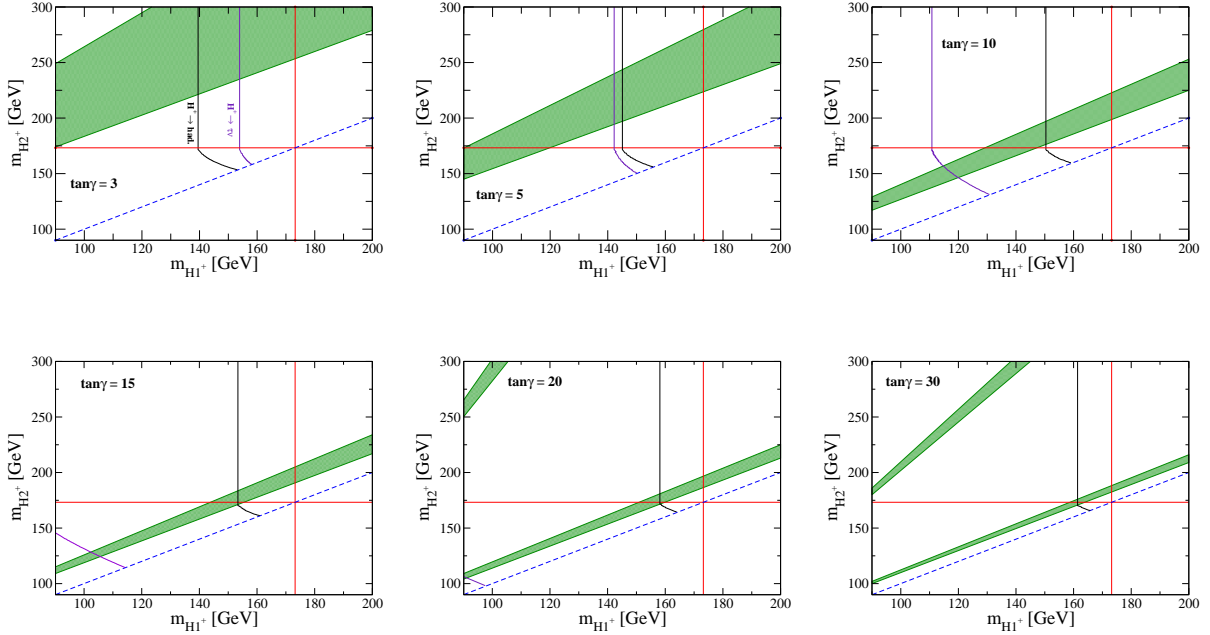

\begin{center}
\includegraphics[width=50mm]{const_tang3.eps}\hspace{3mm}
\includegraphics[width=50mm]{const_tang5.eps}\hspace{3mm}
\includegraphics[width=50mm]{const_tang10.eps} \\\vspace{10mm}
\includegraphics[width=50mm]{const_tang15.eps}\hspace{3mm}
\includegraphics[width=50mm]{const_tang20.eps}\hspace{3mm}
\includegraphics[width=50mm]{const_tang30.eps}
\caption{Green shaded regions indicate the 2$\sigma$ allowed region by the $B\to X_s\gamma$ data in the Type-Y 3HDM with $\tan\beta=2$ and $\theta_C=-\pi/4$. 
The value of $\tan\gamma$ is taken to be 3, 5, 10, 15, 20 and 30 from the upper left to lower right panel. 
The right region from the black and purple curve is allowed by 
the bound from the $t \to H_{1,2}^\pm b \to q\bar{q}' b$ ($cs$ and $cb$) and $t \to H_{1,2}^\pm b \to \tau\nu b$ processes at the LHC. 
The branching ratio of $H_1^\pm \to cb$ is about 43\%, 68\%, 82\%, 84\%, 85\% and 86\% for the case with $\tan\gamma = 3$, 5, 10, 15, 20 and 30, respectively, 
and $m_{H_1^\pm} < m_t$.  
}
\label{const1}
\end{center}
\end{figure}

In Fig.~\ref{const1}, we show the allowed parameter space on the $m_{H_1^\pm}^{}$-$m_{H_2^\pm}^{}$ plane in the 
Type-Y 3HDM with $\tan\beta = 2$ and $\theta_C=-\pi/4$. 
The green shaded region is allowed from  $B\to X_s\gamma$ data 
and the right region from the purple and the black curve satisfies the requirement given in Eqs.~(\ref{leptonic}) and (\ref{had}), respectively. 
The value of $\tan\gamma$ is taken to be 3, 5, 10, 15, 20 and 30 as indicated in each panel of the figure. 
We note that the region below the dashed curve, $m_{H_1^\pm}^{}>m_{H_2^\pm}$, is excluded by definition. 
It is seen that the constraint from $H_{1,2}^\pm \to q\bar{q}'$ becomes stronger as compared to that from $H_{1,2}^\pm \to \tau\nu$ when we take a larger value of $\tan\gamma$, 
because of the enhancement of ${\rm BR}(H_{1,2}^\pm \to cb/cs)$, as we already saw  in Fig.~\ref{br1}. 
Consequently, the case with $m_{H_1^\pm}< m_t -m_b$ and $m_{H_2^\pm}< m_t -m_b$ is highly constrained from $B\to X_s\gamma$ and the direct search at the LHC. 
However, we can find  allowed regions with $m_{H_1^\pm}< m_t -m_b$ and $m_{H_2^\pm} > m_t -m_b$ and also those with
$m_{H_1^\pm} > m_t -m_b$ and $m_{H_2^\pm} > m_t -m_b$. 
The former case is phenomenologically very interesting, because the lighter charged Higgs boson $H_1^\pm$ can mainly decay into the $cb$ final state. 
The BR of the $H_1^\pm \to cb$ mode is shown in the caption of 
Fig.~\ref{const1}, and is essentially determined only by the value of 
$\tan\gamma$ for a fixed value of $\tan\beta$ and $\theta_C$ when 
$m_{H_1^\pm}< m_t$. 

\begin{figure}[t]
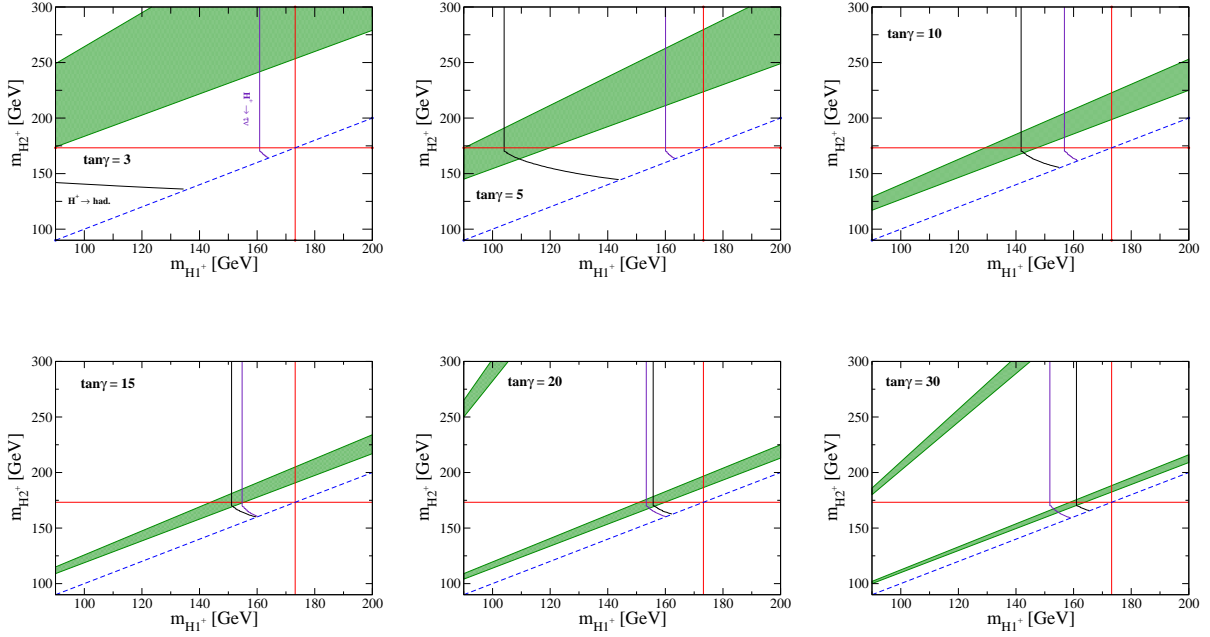

\begin{center}
\includegraphics[width=50mm]{const_tang3_Z.eps}\hspace{3mm}
\includegraphics[width=50mm]{const_tang5_Z.eps}\hspace{3mm}
\includegraphics[width=50mm]{const_tang10_Z.eps} \\\vspace{10mm}
\includegraphics[width=50mm]{const_tang15_Z.eps}\hspace{3mm}
\includegraphics[width=50mm]{const_tang20_Z.eps}\hspace{3mm}
\includegraphics[width=50mm]{const_tang30_Z.eps}
\caption{Same as Fig.~\ref{const1}, but for the case of Type-Z 3HDM. 
The branching ratio of $H_1^\pm \to cb$ is about 4\%, 16\%, 50\%, 67\%, 75\% and 81\% for the case with $\tan\gamma = 3$, 5, 10, 15, 20 and 30, respectively, 
and $m_{H_1^\pm} < m_t$. }
\label{const2}
\end{center}
\end{figure}

In Fig.~\ref{const2}, we also show a similar plot for the Type-Z 3HDM, where the constraint from $B\to X_s\gamma$ is exactly the same as that in the 
Type-Y 3HDM because of 
the same structure of the quark Yukawa couplings. 
The difference can be seen in the relative strength of the constraint from Eqs.~(\ref{leptonic}) and (\ref{had}) as compared to the Type-Y case. 
In analogy with the Type-Y case, the scenario with both charged Higgs boson masses 
smaller than $m_t-m_b$ is highly constrained by  $B\to X_s\gamma$ and LHC direct searches, but 
at least one of the charged Higgs bosons can be lighter than the top quark. 
Similarly to Fig.~\ref{const1}, 
we give the value of the BR of $H_1^\pm \to cb$ in the caption of this figure.

\begin{figure}[t]
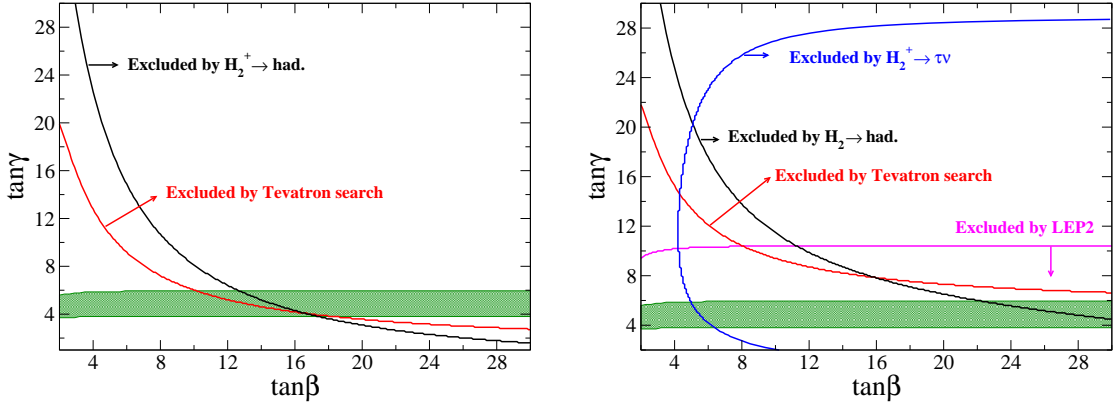

\begin{center}
\includegraphics[width=70mm]{const_83-160.eps}\hspace{6mm}
\includegraphics[width=70mm]{const_83-160_Z.eps}
\caption{Allowed parameter space by several constraints on the $\tan\beta$-$\tan\gamma$ plane 
in the Type-Y (left) and Type-Z (right) 3HDM with $m_{H_1}=83$ GeV, $m_{H_2^\pm}=160$ GeV and $\theta_C=-\pi/4$.
The right region from the blue, black and red curve is excluded by Eq.~(\ref{leptonic}), Eq.~(\ref{had}) and the Tevatron data, respectively. 
The region below the magenta curve is also excluded by the LEP2 data. The green shaded region is allowed by the measurement of $B\to X_s\gamma$.  
}
\label{const3}
\end{center}
\end{figure}

It is particularly interesting to investigate the case with a charged Higgs boson mass  between 80 GeV and 90 GeV, and $H^\pm$ decaying dominantly to $cs/cb$.         
In this situation, the bound from  direct searches at the LHC cannot be applied to exclude any parameter space 
because  no sensitivity exists in this mass region. This is because the background from $t\to W^\pm b$
is overwhelming in this region and the invariant mass cut on the jets originating from $H^\pm$ would 
lose its effect of greatly suppressing the background when $m_{H^\pm}$ is close to $m_W$. }
However, searches for $H^\pm$ from  LEP2 and the Tevatron have some sensitivity to this region of charged Higgs boson mass  between 80 GeV and 90 GeV
and we include these bounds in our analysis.
In Ref.~\cite{LEP2}, the excluded region in the $m_{H^\pm}$-BR($H^\pm \to \tau\nu$) plane has been given by using the combined LEP2 data
from all four experiments.
If we take $m_{H_1^\pm}=83$ GeV, we can extract the bound BR$(H_1^\pm \to \tau\nu)\lesssim 0.45$ at the 95\% CL, with
the exact bound depending on the choice of $m_{H^\pm_1}$ i.e. there is a sizeable
region of unexcluded parameter space where BR$(H_1^\pm \to cs/cb)$ is sizeable and $80\, {\rm GeV} < m_{H_1^\pm}< 90$ GeV.
 We note that the results of the LEP2 search in Ref.~\cite{LEP2}
show that there are some regions of  BR$(H_1^\pm \to cs/cb)$ and $80\,{\rm GeV }< m_{H^\pm_1}< 90$ GeV 
where there are fluctuations in excess of $2\sigma$ above the background.
In Ref.~\cite{Tevatron}, we can also extract the bound BR$(t \to H^+ b) \times \text{BR}(H_1^\pm \to q\bar{q}')\lesssim 0.2$  at the 95\% CL
from the data collected at the Tevatron with 1.0 fb$^{-1}$. The reason why this Tevatron search has sensitivity to 
charged Higgs masses between 80 GeV and 90 GeV is because no invariant mass cut is used, and instead a disappearance search is carried out. 
So far the
LHC searches have not used this search strategy.

By imposing these two constraints (from LEP2 and Tevatron) for $H_1^\pm$ 
and those from Eqs.~(\ref{leptonic}) and (\ref{had}) for $H_2^\pm$, 
we obtain the excluded region on the $\tan\beta$-$\tan\gamma$ plane  shown in Fig.~\ref{const3}. 
In this figure, we take $m_{H_1^\pm}=83$ GeV, $m_{H_2^\pm}=160$ GeV and $\theta_C = -\pi/4$. 
The left and right panel show the case in the Type-Y and Type-Z 3HDM, respectively. 
The green shaded region is allowed by the $B\to X_s\gamma$  data. 
As we can see,  in the Type-Z 3HDM, there is no region satisfying all the constraints mentioned above. 
In contrast, we can find  allowed regions in the Type-Y case, namely, when $4\lesssim \tan\gamma \lesssim 6$ and 
$\tan\beta <10\text{-}18$.

\begin{table}[t]
\begin{center}
\begin{tabular}{cc||c|c|c|cc}\hline\hline
&$(m_{H_1^\pm},m_{H_2^\pm},\tan\gamma)$ & ${\rm BR}(H_1^\pm \to X)_{\text{Y}}$  & ${\rm BR}(H_2^\pm \to X)_{\text{Y}}$ & ${\rm BR}(H_1^\pm \to X)_{\text{Z}}$  & ${\rm BR}(H_2^\pm \to X)_{\text{Z}}$          \\  \hline\hline
BM1:&$(83,160,5)$&   $cb:68,cs:17$ &$cb:77,cb:16$  &- &-   \\\hline
BM2:&$(160,250,3)$&   $cb:43,\tau\nu:34$ &$tb:99.7,cb:0.17$  &$\tau\nu:94,cb:4.0$ &$tb:99.7,ts:0.17$   \\\hline
BM3:&$(160,225,5)$&   $cb:68,cs:17$&   $tb:99.6,cb:0.21$&$\tau\nu:80,cb:16$&$tb:99.4,cb:0.21$ \\\hline
BM4:&$(160,200,10)$&  $cb:82,cs:15$&  $tb:98,cb:1.3$&$cb:50,\tau\nu:41$&$tb:98,cb:1.3$\\\hline
BM5:&$(160,180,20)$&  $cb:85,cs:14$&  $tb:67,cb:28$&$cb:75,\tau\nu:13$  &$tb:66,cb:27$ \\\hline
BM6:&$(200,250,10)$&  $tb:99.0,cb:0.89$&  $tb:99.5,cb:0.29$&$tb:98,cb:0.89$  &$tb:99.4,cb:0.29$ \\\hline\hline
\end{tabular} 
\end{center}
\caption{Predictions of the biggest two values of ${\rm BR}(H_{1,2}^\pm \to X)$ in the Type-Y and Type-Z 3HDM for the six benchmark points (BM1-BM6) which are allowed 
by $B\to X_s\gamma$ and direct searches at LEP2, Tevatron and LHC. 
We take $\tan\beta =2$ and $\theta_C=-\pi/4$. 
The values of $m_{H_1^\pm}$ and $m_{H_2^\pm}$ are presented in GeV while those for the BRs are in \%. 
For BM1, only the Type-Y is allowed, so that we do not show the predictions in the Type-Z 3HDM. 
}
\label{pred}
\end{table}

\section{Phenomenology of 3HDM charged Higgs bosons at the LHC}

\begin{figure}[t]
\begin{center}
\includegraphics[width=150mm]{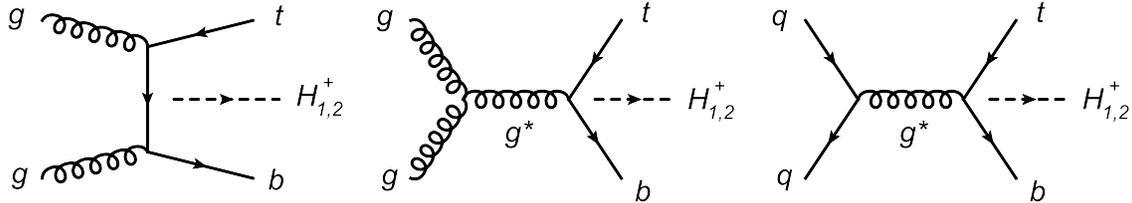}
\caption{Topologies of the Feynman diagram for the $gg\to \bar{t} b H^+_{1,2}$ (left and middle) 
and $q\bar{q} \to \bar{t} b H^+_{1,2}$  (right) processes. 
The charged Higgs bosons can be emitted from the final state quark current at 3 (left) and 2 (middle and right) different points. 
For the topology on the left, gluon permutations are also required. In total, one has 8(2) diagrams for the
$gg(q\bar{q})$-induced process (at fixed $q$ flavour).}
\label{diag}
\end{center}
\end{figure}

The collider phenomenology of a charged Higgs boson can be classified into two regimes depending 
on its mass $m_{H^\pm}^{}$: (i)
$m_{H^\pm} <m_t-m_b$ (light) and (ii) $m_{H^\pm} > m_t-m_b$ (heavy). 
For  case (i), charged Higgs bosons can be produced via the top quark decay, so that the 
main production process at the LHC is $gg,q\bar q \to t\bar{t} \to H^+ b\, \bar{t}$ (see Fig.~\ref{diag}). 
For  case (ii), the main production mode is the top quark associated process, i.e., $gb \to H^\pm t$ + c.c. 
As intimated, See Fig.~\ref{diag}, we will use the  $gg\to t\bar{b}H^-$ + c.c. subprocess (at LO) 
in our Monte Carlo (MC) analysis, which captures both (i) (limited to the $gg$ channel) and (ii) as well as
their interference, which is important in the threshold region
$m_{H^\pm} \sim m_t-m_b$~\cite{Guchait:2001pi,Assamagan:2004gv}.
(In fact, we will also be emulating the subleading contribution from $q\bar{q}\to t\bar b H^-$ + c.c.)
Recall that, on the one hand, in the narrow width approximation of the top quark one has that
$\sigma(gg,q\bar{q}\to t\bar{b} H^-) \equiv \sigma(gg,\, q\bar{q}\to t\bar{t})\times{\rm BR}(\bar{t}\to \bar{b} H^-)$ 
(limited to the diagrams in which the $H^-$ is emitted by the $t$ antiquark) 
and, on the other hand, the $b$-quark in the initial state comes from a gluon splitting inside the proton, 
as explained in \cite{Guchait:2001pi,Assamagan:2004gv}.

In order to encourage  phenomenological studies for the charged Higgs bosons at the LHC, 
we present  six benchmark parameter sets, BM1--BM6, allowed by  the $B\to X_s\gamma$ data and direct searches at LEP2, Tevatron and LHC in Tab.~\ref{pred}. 
Herein, the biggest two values of BRs for $H_1^\pm$ and $H_2^\pm$ are given in the Type-Y and Type-Z 3HDM for each of the six benchmark points. 
We will particularly use BM1, BM4 and BM6 in the Type-Y case for our forthcoming MC analysis, 
as illustrative of the three situations emerged so far: of a light, mixed and heavy charged Higgs mass spectrum, respectively, with respect to the top quark mass.
While we refer to the Type-Y 3HDM case in the remainder of our analysis, 
we confirm that the ensuing phenomenology is not dissimilar in the Type-Z 3HDM case (except for BM1 which is not allowed herein).

For the MC study we have computed the following signal ($S$) and (irreducible) background ($B$) processes, respectively:
\begin{align}
&\text{a.}~ gg,q\bar{q}\to t\bar{b} H^-_{1,2} + \text{c.c}  \to t\bar b  jj + \text{c.c}, \notag\\
&\text{b.}~ gg,q\bar{q}\to t\bar{b} W^- + \text{c.c.} \to t\bar{b}  jj + \text{c.c.}, \notag
\end{align}
where the di-jet system $jj$ is tagged through a single $b$-tag, as recommended in \cite{Akeroyd2}. We remind the reader here that 
applying such a {$b$}-tag would improve sensitivity to $H^\pm_{1,2} \to cb$ decays greatly, as  the background from 
{$W\to cb$} has a very small rate. This is made explicit by choosing a
  {$b$}-tagging efficiency {$\epsilon_b=0.5$}, a {$c$}-quark mistagging rate  {$\epsilon_c=0.1$} and a
 light quark $(u,d,s)$ mistagging rate 
{$\epsilon_j=0.01$}. It follows  that the
 estimate gain in sensitivity with respect to the case in which the di-jet system is untagged is then:
\begin{equation}
\frac{[S/\sqrt B]_{\rm btag}}{[S/\sqrt B]_{\rm \not{btag}}}
\sim \frac{\epsilon_b\sqrt 2}{\sqrt{(\epsilon_j+\epsilon_c)}}\sim {2.13}.
\end{equation}

\begin{figure}[t]
\begin{center}
\includegraphics[width=50mm,angle=90]{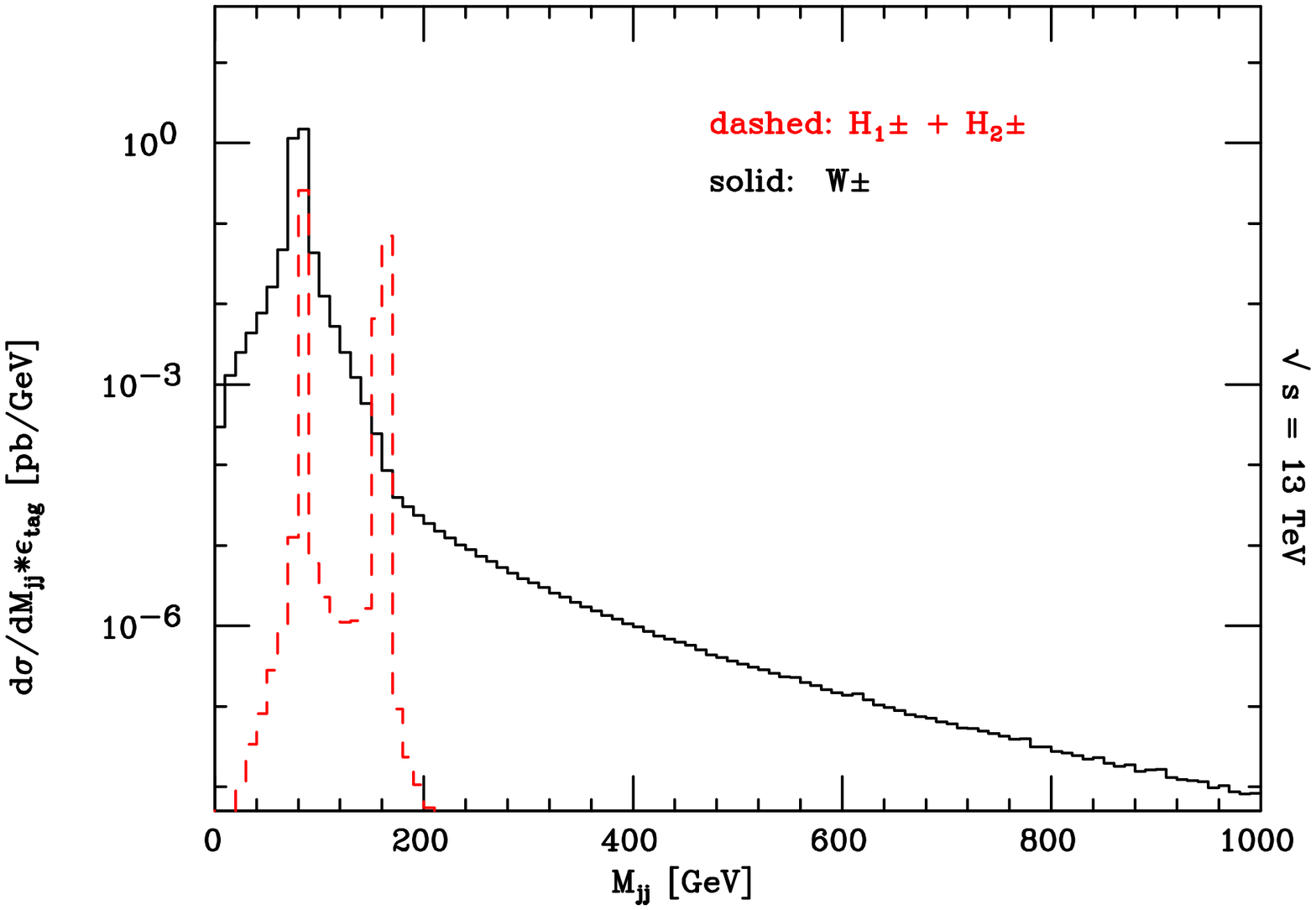}\hspace{3mm}
\includegraphics[width=50mm,angle=90]{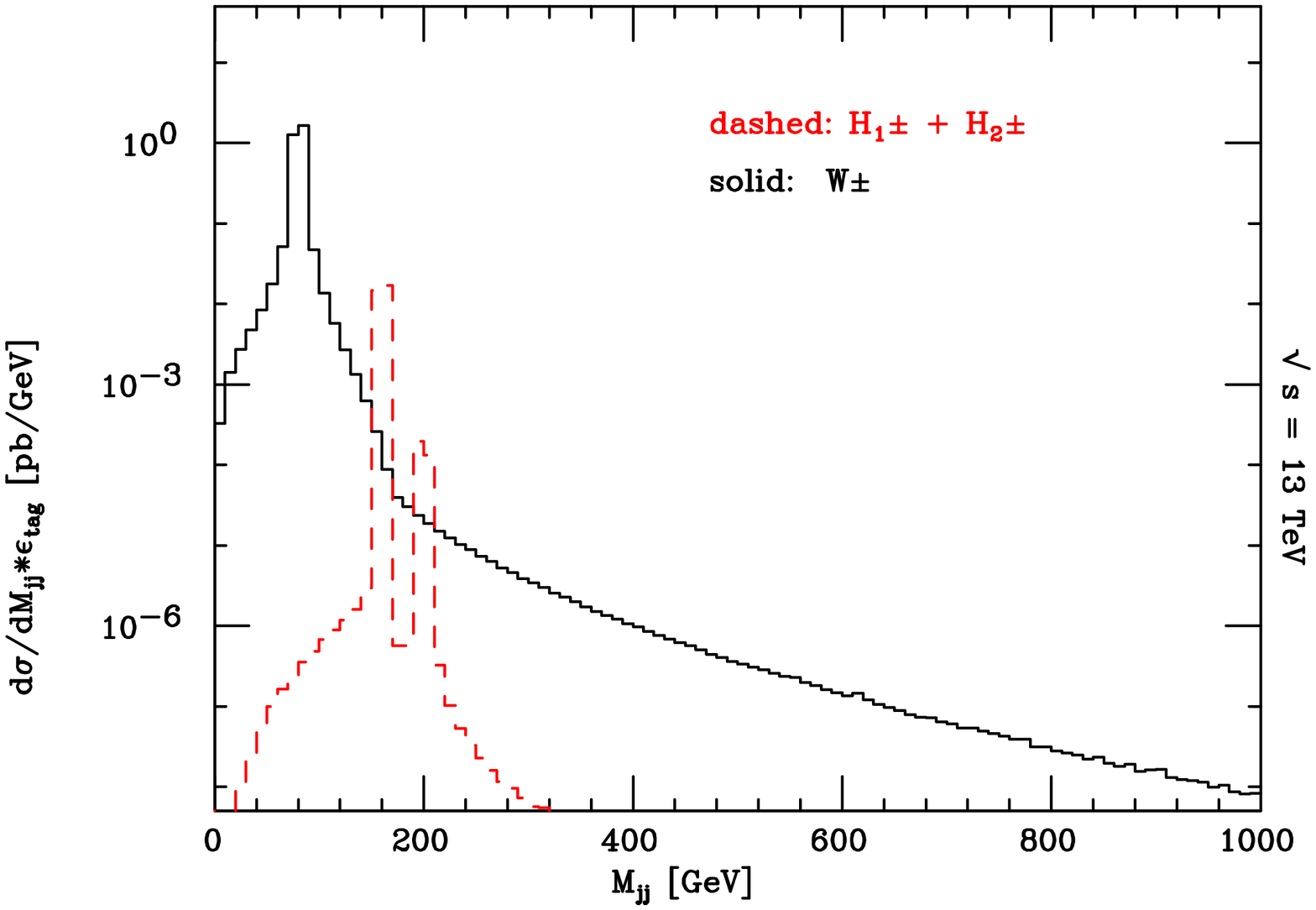}\vspace{5mm}
\includegraphics[width=50mm,angle=90]{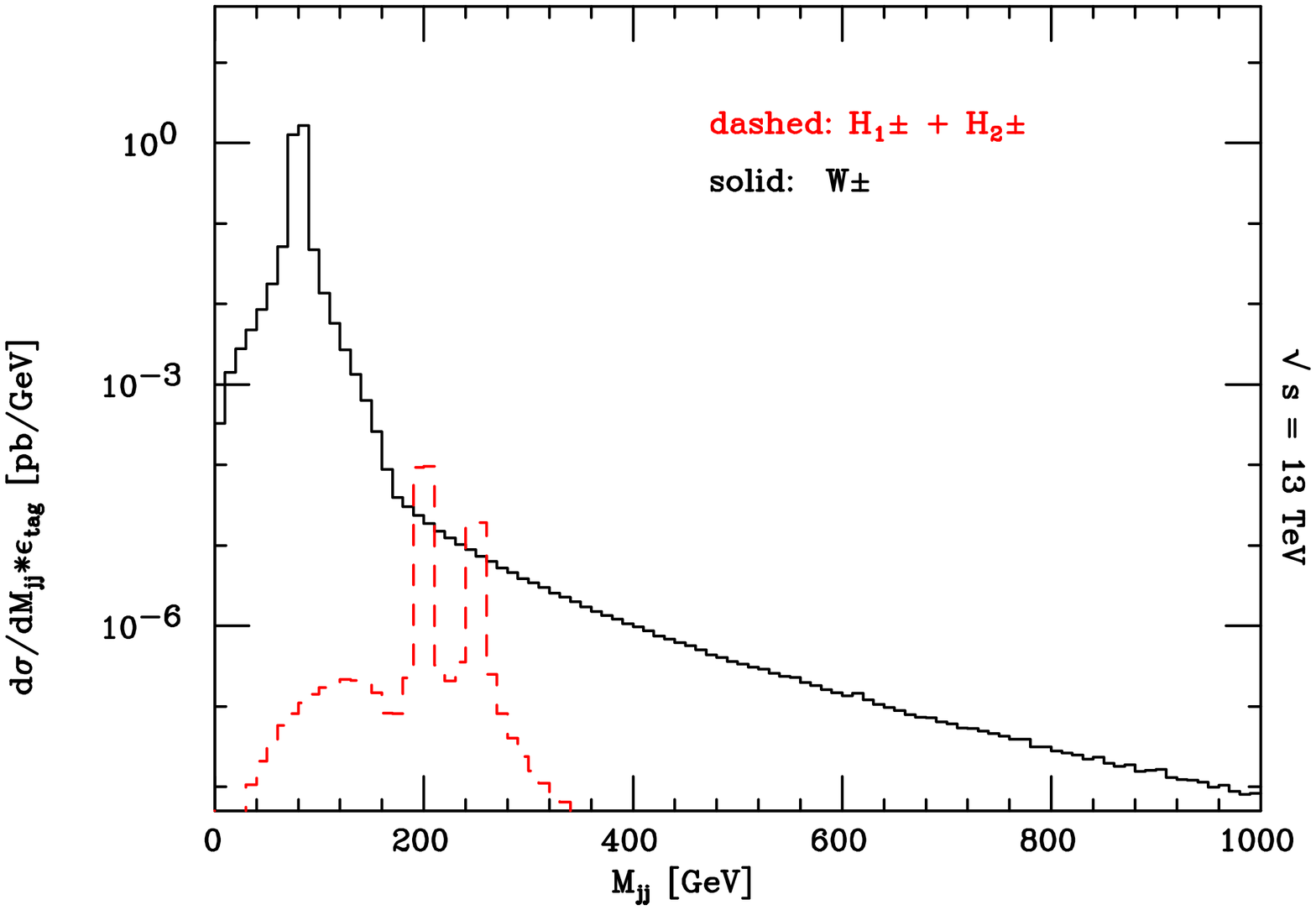} 
\caption{Differential distributions in the di-jet invariant mass of  processes $a$ (red dashed) and $b$ (black solid) 
for the BM1 (top-left), BM4 (top-right) and BM6 (bottom) at the LHC with $\sqrt s=13$ TeV. Tagging efficiencies are included as described in the text.
CTEQ(4L) with $Q=\mu=\sqrt{\hat s}$ is used \cite{Lai:1996mg}. }
\label{fig:twopeaks}
\end{center}
\end{figure}

Fig.~\ref{fig:twopeaks} shows the di-jet mass distribution for $S$ and $B$  at 13 TeV in terms of cross section for BM1, BM4 and BM6. 
Even before enforcing any selection cuts, it is clear the
LHC potential in accessing these peculiar 3HDM signatures during Run 2. Two caveats should be borne in mind here though.
On the one hand, we have not allowed for full combinatorial effects in the di-jet mass reconstruction, as we have assumed
that each of the three $b$-jets present in the final state can be correctly assigned to its parent heavy particle (i.e., $t$,
$\bar t$ and $H^\pm_{1,2}$). On the other hand, BM1, BM4 and BM6 are the very
best points for our purposes, those with highest BR, while one really ought to test the entire parameter space of 3HDMs sampled  over the 
inputs  $\theta_C$, $\tan\beta$, $\tan\gamma$, $m_{H^\pm_1}$ and $m_{H^\pm_2}$.
Nonetheless, we believe that the very peculiar $H^\pm_{1,2}$ mass patterns that we have discussed deserve further
investigation in presence of parton shower, hadronisation, jet reconstruction and detector effects \cite{preparation}.

Finally, we briefly comment on the phenomenology of the additional neutral Higgs bosons $H_{1,2}$ and $A_{1,2}$ in the 3HDMs. 
When we consider the case with the masses of $H_{1,2}^\pm$ to be ${\cal O}(100)$ GeV, the 
neutral Higgs bosons cannot be so heavy due to the constraints from 
electroweak precision observables such as the $S$ and $T$ parameters~\cite{hernandez} 
and from perturbative unitarity~\cite{3hdm-uni}. If we consider the case where these neutral Higgs bosons are heavier than 
the charged Higgs bosons, then there are no additional decay modes
of the charged Higgs bosons beyond those shown in this paper. 
However, in this case, the decay channels 
$H_{1,2}/A_{1,2}\to H_{1,2}^\pm W^{(*)^\mp}$ can be dominant depending 
on the mass difference between the
neutral Higgs states and the charged Higgs states, and the 
numerical values of the $X_a$, $Y_a$ and $Z_a$ parameters defined 
in Eqs.~(\ref{yuk2a})-(\ref{xyz2}). 
These cascade decay channels would be additional production modes of
$H_{1,2}^\pm$ beyond those studied in this section, although they 
would require a separate signal-background study in order to 
assess the detection prospects.

\section{Conclusions}

We have discussed the phenomenology of  charged Higgs bosons in fully active 3HDMs with two softly-broken discrete $Z_2$ symmetries which are imposed to 
avoid  FCNCs at  tree level. 
Under these $Z_2$ symmetries, we have defined five types of Yukawa interactions.
We have then shown that the decay branching fractions of 
$H_{1,2}^\pm \to cb$ can be dominant in the Type-Y and Type-Z 3HDM when the masses of the charged Higgs bosons
are taken to be below $m_t-m_b$. 
The $H^\pm \to cb$ decay can also be dominant in the Type-Y 2HDM with $m_{H^\pm}< m_t-m_b$, but such a light charged Higgs boson scenario 
is excluded by the constraint from $B\to X_s\gamma$. In contrast, in 3HDMs,
the scenario with masses of ${\cal O}(100)$ GeV for the charged Higgs bosons is allowed by $B\to X_s\gamma$
because of a cancellation between the separate contributions from the $H_1^\pm$ and $H_2^\pm$ loop diagrams. 
Therefore, the search for a light charged Higgs boson decaying into $cb$ is a  means to distinguish  3HDMs from  2HDMs. 

We then have calculated the branching fraction of the $B\to X_s\gamma$ process at NLO in QCD in 3HDMs in order to confirm how the 
cancellation takes place numerically. 
We found that it happens especially 
in the Type-II, Type-Y and Type-Z 3HDMs 
when there is a non-zero mixing and a mass difference between $H_1^\pm$ and $H_2^\pm$. 
In the Type-I and Type-X 3HDMs, the numerical values of ${\rm BR}(B\to X_s\gamma)$ are not much different from the predictions in  Type-I and Type-X 2HDMs. 

We also have taken into account the constraints from direct searches at the LHC of charged Higgs bosons from the top quark decays
 $t \to H^\pm b \to \tau\nu b$ and 
$t \to H^\pm b \to q\bar{q}' b$ with the 8 TeV data. 
We have found that, in the Type-Y and Type-Z 3HDM, 
the scenario with both $H_1^\pm$ and $H_2^\pm$ lighter than $m_t-m_b$ 
is highly constrained from $B\to X_s\gamma$ and the LHC direct searches, while 
the scenario with only $H_1^\pm$ lighter than $m_t-m_b$ is allowed. 
However, the particular case $m_{H^\pm_1}\approx m_{W^\pm}$  with $m_{H^\pm_2}< m_t$ is allowed (also by Tevatron and LEP2), albeit only in the 
Type-Y 3HDM. We drew attention to the fact that the region of $80\,{\rm GeV}< m_{H^\pm_1}< 90$ GeV is not constrained
by current LHC searches for $t\to H^\pm b$ followed by dominant decay $H^\pm\to cs/cb$, and this parameter
space is only weakly constrained from LEP2 and Tevatron searches. Any future signal in this region could be
readily accommodated by $H^\pm_1$ from a 3HDM

Finally, upon running a MC simulation to compare the yield of the $H^\pm_{1,2}$ signals and $W^\pm$ background 
through
the production processes $gg,q\bar q\to t\bar b H^-_{1,2}$ + c.c   and
$gg,q\bar q\to t\bar b W^-$ + c.c, respectively, followed by the corresponding 
di-jet decays $H^\pm_{1,2}\to jj$ and 
$W^\pm\to jj$, we have shown that  the aforementioned charged Higgs boson signals should be accessible
at Run 2 of the LHC over a suitable region of the 3HDM parameter space,
provided that $b$-tagging is enforced so as to single out the $cb$ component above the $cs$ one.
Therefore, these (multiple) charged Higgs boson  signatures can be used not only to distinguish between 2HDMs and 
3HDMs but also to identify the types realising the latter.

\section*{Acknowledgements}
\noindent
K.Y. is supported by a JSPS Postdoctoral Fellowships for Research Abroad.
S.M. is financed in part through the NExT Institute. 
E.Y. is supported by the Ministry of  National Education of Turkey.

\section*{Note added:}
\noindent
After this work was completed an explicit search for $t\to H^+ b$ 
followed by the decay $H^+ \to cb$ 
was carried out by the CMS collaboration in CMS PAS HIG-16-030, with 19.7 fb$^{-1}$ of data at $\sqrt{s}= 8$ TeV. 
Stronger upper limits on the branching ratio of
$t\to  H^+b$ have been probed than for the CMS search for $t\to H^+b$, $H^+\to cs$ in Ref.~\cite{hadronic} with the same data sample. 
This result explicitly shows the increase in sensitivity that can be obtained by tagging a third $b$ quark.

\begin{appendix}

\newpage 
\section{Mass matrix elements}

We present  here the analytic expressions for the mass matrices of the singly-charged 
(${\cal M}_C^2$), CP-odd (${\cal M}_A^2$) and CP-even (${\cal M}_H^2$) scalar states in the Higgs basis of 3HDMs. 
The matrix elements of ${\cal M}_C^2$ and  ${\cal M}_A^2$ are given by 
\begin{align}
({\cal M}_C^2)_{11} &= -\frac{v^2}{2}\left[(\rho_2 + \rho_3)c_\gamma^2 + (\kappa_2 + \kappa_3)s_\gamma^2)   \right] 
+ \mu_{12}^2\frac{c_\gamma}{s_\beta c_\beta}+\mu_{23}^2\frac{s_\gamma}{s_\beta c_\beta}  , \\
({\cal M}_C^2)_{22} &= -\frac{v^2}{2}\left[(\rho_2 + \rho_3)s_\beta^2s_\gamma^2 + (\sigma_2 + \sigma_3)c_\beta^2 + (\kappa_2 + \kappa_3)s_\beta^2c_\gamma^2   \right]\notag\\
&\quad +\mu_{12}^2t_\beta s_\gamma t_\gamma + \frac{\mu_{13}^2}{s_\gamma c_\gamma} +\mu_{23}^2 \frac{t_\beta c_\gamma}{ t_\gamma} , \\
({\cal M}_C^2)_{12} &= -\frac{v^2}{4}(\rho_2 + \rho_3 -\kappa_2 -\kappa_3)s_\beta s_{2\gamma}  
+ \mu_{12}^2\frac{s_\gamma}{c_\beta}-\mu_{23}^2\frac{c_\gamma}{c_\beta}  , \\
({\cal M}_A^2)_{11} &= -v^2\left(\rho_3 c_\gamma^2 +\kappa_3 s_\gamma^2\right) +  \mu_{13}^2\frac{c_\gamma}{s_\beta c_\beta} + \mu_{23}^2\frac{s_\gamma}{s_\beta c_\beta}, \\ 
({\cal M}_A^2)_{22} &= -v^2\left(\rho_3 s_\beta^2s_\gamma^2 +\sigma_3 c_\beta^2 +\kappa_3 s_\beta^2 c_\gamma^2 \right) 
+\mu_{12}^2 t_\beta s_\gamma t_\gamma + \mu_{13}^2 \frac{s_\gamma}{s_\gamma c_\gamma} + \mu_{23}^2 \frac{t_\beta c_\gamma}{t_\gamma} , \\ 
({\cal M}_A^2)_{12} &= -\frac{v^2}{2}(\rho_3-\kappa_3)s_\beta s_{2\gamma} + \mu_{12}^2\frac{s_\gamma}{c_\beta} - \mu_{23}^2\frac{c_\gamma}{c_\beta}.
\end{align}
Those for ${\cal M}_H^2$ are given by 
\begin{align}
({\cal M}_H^2)_{11} &= \frac{v^2}{2}(2\lambda_1c_\beta^4c_\gamma^4 + 2\lambda_2s_\beta^4 + 2\lambda_3c_\beta^4s_\gamma^4 + 
 \rho_{123}s_{2\beta}^2 c_\gamma^2 + \sigma_{123}c_{\beta}^4 s_{2\gamma}^2 +\kappa_{123}s_{2\beta}^2s_\gamma^2), \\
({\cal M}_H^2)_{22} &= \frac{v^2}{8}(2\lambda_1 s_{2\beta}^2c_\gamma^4 + 2\lambda_2 s_{2\beta}^2 + 2\lambda_3s_{2\beta}^2s_\gamma^4 
-4\rho_{123} c_\beta^2 s_{2\gamma}^2  + \sigma_{123} s_{2\beta}^2s_{2\gamma}^2  - 4\kappa_{123} s_{2\beta}^2s_\gamma^2  )\notag\\
&\quad +\mu_{12}^2\frac{c_\gamma}{s_\beta c_\beta} + \mu_{23}^2\frac{s_\gamma}{s_\beta c_\beta},\\
({\cal M}_H^2)_{33} &= \frac{v^2}{4}(\lambda_1 + \lambda_3 -2\sigma_{123})c_\beta^2s_{2\gamma}^2  
+\mu_{12}^2t_\beta s_\gamma t_\gamma + \frac{\mu_{13}^2}{s_\gamma c_\gamma} + \mu_{23}^2 \frac{c_\gamma t_\beta}{t_\gamma}, \\
({\cal M}_H^2)_{12} & = -\frac{v^2}{2}\Big[
2\lambda_1 s_\beta c_\beta^3 c_\gamma^4 - 2\lambda_2 s_\beta^3 c_\beta + 2\lambda_3 s_\beta c_\beta^3 s_\gamma^4\notag\\
&\quad\quad\quad-(c_\beta + c_{3\beta})s_\beta c_\gamma^2 \rho_{123} + s_\beta c_\beta^3 s_{2\gamma}^2\sigma_{123} - (c_\beta + c_{3\beta})s_\beta s_\gamma^2 \kappa_{123}\Big], \\
({\cal M}_H^2)_{13} & = -\frac{v^2}{4}\left(
2\lambda_1c_\beta^3 s_{2\gamma} c_\gamma^2 - 2\lambda_3 c_\beta^3s_{2\gamma}s_\gamma^2 
+ 2\rho_{123} s_\beta^2c_\beta  s_{2\gamma} - \sigma_{123}c_\beta^3s_{4\gamma} - 2\kappa_{123}s_\beta^2 c_\beta s_{2\gamma}  \right), \\
({\cal M}_H^2)_{23} & = \frac{v^2}{4}[
4\lambda_1  s_\gamma c_\gamma^3 - 4\lambda_3  s_\gamma^3 c_\gamma
-2(\rho_{123}-\kappa_{123})s_{2\gamma}-\sigma_{123} s_{4\gamma} ]s_\beta c_\beta^2
+ \mu_{12}^2\frac{s_\gamma}{c_\beta}-\mu_{23}^2\frac{c_\gamma}{c_\beta}. 
\end{align}
In the above expressions, we used the shorthand notations $c_X^{}=\cos X$, $s_X^{}=\sin X$ and $t_X = \tan X$.

\section{Input parameters}

For the numerical evaluations, we have used the following input values for the SM parameters~\cite{PDG}:
\begin{align}
&m_{W^\pm}= 80.385~\text{GeV},~m_Z=91.1876~\text{GeV},~G_F=1.1663787\times 10^{-5}~\text{GeV}^{-2},\notag\\
&m_\tau=1.77684~\text{GeV},~m_\mu= 0.105658367~\text{GeV}, \notag\\
&\alpha_s(m_Z^{})=0.1185,~m_t^{\text{pole}} = 174.6~\text{GeV},~m_b^{\text{pole}}= 4.89~\text{GeV},~m_c^{\text{pole}}= 1.64~\text{GeV}, 
\end{align}
where $m_t^{\text{pole}}$, $m_b^{\text{pole}}$ and $m_c^{\text{pole}}$ are respectively the pole masses of the top, bottom and charm quark. 
For the $\overline{\text{MS}}$ masses of the quarks, we use the following values~\cite{PDG}:
\begin{align}
\bar{m}_b(\bar{m}_b) = 4.18~\text{GeV},~
\bar{m}_c(\bar{m}_c) = 1.275~\text{GeV},~
\bar{m}_s(2~\text{GeV}) = 0.0935~\text{GeV}.
\end{align}
Using these $\overline{\text{MS}}$ masses, we obtain the running masses of the quarks at, e.g., $\mu = 100$ GeV to be 
$\bar{m}_b=3.01$ GeV, $\bar{m}_c=0.701$ GeV and $\bar{m}_s=0.0489$ GeV.
As for the other inputs, we use 
\begin{align}
|V_{tb}V_{ts}/V_{cb}|^2 = 0.9626,~{\rm BR}(B \to X_c\ell\nu)= 0.1065. 
\end{align}
for the calculation of ${\rm BR}(B\to X_s\gamma)$ and 
\begin{align}
|V_{cb}| = 0.0409 
\end{align}
for that of the charged Higgs boson decays.

\end{appendix}

\end{document}